\documentclass[letterpaper,nofootinbib,floatfix,superscriptaddress,preprintnumbers]{revtex4}

\usepackage{color}
\usepackage{array}
\usepackage{graphicx}
\usepackage{hyperref}
\usepackage{tikz}
\tikzset{>=latex}
\usetikzlibrary{trees}
\usetikzlibrary{decorations.pathmorphing}
\usetikzlibrary{decorations.markings}
\usetikzlibrary{decorations.pathreplacing}
\tikzset{
    gluon/.style={decorate, draw=black,
        decoration={coil,amplitude=3pt, segment length=4pt,aspect=0.7}} 
}
\tikzset{
    photon/.style={decorate, decoration={snake}},
}
\tikzset{
    photonvalley/.style={decorate, draw=black,
        decoration={snake,amplitude=2pt, segment length=6pt,aspect=1.7}} 
}

\hypersetup{
  colorlinks   = true, 
  urlcolor     = blue, 
  linkcolor    = blue, 
  citecolor   = red 
}

\usepackage{verbatim}
\usepackage{amsmath}
\usepackage{amssymb}
\usepackage{subfigure}
\usepackage{url}
\usepackage{array}
\usepackage{slashed}

\usepackage{multirow}
\usepackage{threeparttable}
\usepackage{paralist}

\usepackage{color}
\definecolor{darkgreen}{rgb}{0,0.5,0}

\newcommand{\sig}{\sigma}
\newcommand{\gam}{\gamma}
\newcommand{\ga}{\gamma}

\newcommand{\be}{\begin{equation}}
\newcommand{\ee}{\end{equation}}
\newcommand{\bea}{\begin{eqnarray}}
\newcommand{\eea}{\end{eqnarray}}

\usepackage{soul}

\begin{document}

\title{Rays of light from the LHC}

\author{Simon Knapen} \email{smknapen@lbl.gov}
  \affiliation{Berkeley Center for Theoretical Physics,
University of California, Berkeley, CA 94720}
\affiliation{Theoretical Physics Group, Lawrence Berkeley National Laboratory, Berkeley, CA 94720
}

\author{Tom Melia} \email{tmelia@lbl.gov}
  \affiliation{Berkeley Center for Theoretical Physics,
University of California, Berkeley, CA 94720}
\affiliation{Theoretical Physics Group, Lawrence Berkeley National Laboratory, Berkeley, CA 94720
}

\author{Michele Papucci} \email{mpapucci@lbl.gov}
  \affiliation{Berkeley Center for Theoretical Physics,
University of California, Berkeley, CA 94720}
\affiliation{Theoretical Physics Group, Lawrence Berkeley National Laboratory, Berkeley, CA 94720
}

\author{Kathryn M. Zurek} \email{kzurek@berkeley.edu}
  \affiliation{Berkeley Center for Theoretical Physics,
University of California, Berkeley, CA 94720}
\affiliation{Theoretical Physics Group, Lawrence Berkeley National Laboratory, Berkeley, CA 94720
}

\date{\today}

\preprint{UCB-PTH-15/17}

\begin{abstract}
We consider models for the di-photon resonance observed at ATLAS (with 3.6 fb$^{-1}$) and CMS (with 2.6 fb$^{-1}$).  We find there is no conflict between the signal reported at 13 TeV, and the constraints from both experiments at 8 TeV with 20.3 fb$^{-1}$.  We make a simple argument for why adding only one new resonance to the standard model (SM) is not sufficient to explain the observation.   We explore four viable options: {\em (i)}: resonance production and decay through loops of messenger fermions or scalars; {\em (ii)}: a resonant messenger which decays to the di-photon resonance + X; {\em (iii)}: an edge configuration where $A \rightarrow B \ga \rightarrow C\ga\ga$, and {\em (iv)}: Hidden Valley-like models where the resonance decays to a pair of very light (sub-GeV) states, each of which in turn decays to a pair of collimated photons that cannot be distinguished from a single photon.   Since in each case multiple new states have been introduced, a wealth of signatures is expected to ensue at Run-2 of LHC.
\end{abstract}

\maketitle
\tableofcontents
\clearpage

\section{Introduction}
\label{sec:intro}

Recently ATLAS and CMS have reported an excess in the di-photon mass spectrum around 750~GeV in their first set of data collected at the LHC at 13~TeV. The quoted local significance is $3.6\div3.9\sigma$~\cite{ATLAS-CONF-2015-081} and $2.6\div2.0\sigma$~\cite{CMS-PAS-EXO-15-004} respectively, where the range stems from the variation of the resonance width from narrow to wider, $\Gamma/M = {\cal O}(6\%)$. The discriminating power between small and large width is not very significant and while ATLAS data prefers larger widths, up to 45~GeV, CMS results are more significant in the narrow width approximation. Furthermore ATLAS best fit is for a mass of $747\div750$~GeV, while CMS data peaks at $760$~GeV. In this paper we will assume a value of $750$~GeV for simplicity.
The cross section times branching ratio for the putative signal is consistent with roughly $5\div10\,{\rm fb}$. In the following we will keep the cross section free in this range, except when a single reference value is needed, for which we will assume  $5$ fb, the somewhat conservative lower end of this range. While the likelihood that this excess may still be a statistical fluctuation or some experimental systematic is not negligible, this may well be the first clear signal of physics beyond the standard model and therefore it is worth exploring whether such an effect can be accommodated within some model, and how complicated that model has to be. This exercise can assess the credibility of a new physics (NP) signal and provide further experimental avenues to investigate the excess. 
Before embarking in this activity we should set this new result in perspective against similar 8 TeV searches. Both CMS and ATLAS have set exclusion limits on di-photon resonances at 750~GeV in Run-I~\cite{Khachatryan:2015qba,Aad:2015mna}:
\begin{align*}
&{\rm CMS:} \quad\quad\;\;\;\;1.37 \div 2.41 \,\textrm{fb}\; (0.7 \div 2.0 \,\textrm{fb exp.})&\rightarrow&\quad\quad 6.42 \div 11.3 \,\textrm{fb}\; ( 3.3 \div 9.47\,\textrm{fb exp.})    \\
&{\rm ATLAS:}\quad\quad 2.42 \,\textrm{fb}\; (1.92 \,\textrm{ fb exp.})&\rightarrow&\quad\quad 11.36\,\textrm{fb}\; (9.01 \,\textrm{fb exp.}) 
\end{align*}
where the arrows indicate the rescaled limits to 13 TeV. The range quoted for the CMS analysis corresponds to varying the resonance width in the range $0.1\div75$~GeV, while the ATLAS numbers have been extracted from a public plot and correspond to efficiencies and acceptance calculated for a Randall-Sundrum graviton resonance. We have rescaled the 8~TeV results using the parton luminosity ratio of $4.693$~\cite{higgsxswg} between 8 and 13~TeV, assuming a production mechanism dominated by gluon fusion. In the case of a resonance coupled to $q \bar q$ the limit on the cross section will be a factor of $2.692/4.693=0.574$ smaller.  As one can see the results are broadly consistent across the two runs and both experiments had slight excesses in Run I at the same invariant mass. A more detailed statistical analysis of the consistency of the two results is beyond the scope of this paper.

Clearly, the di-photon final state restricts the spin possibilities to 0 and 2;  in the case of a spin-2, theoretical consistency arguments generally require additional states with masses not too far from the new resonance. Thus we focus most of our discussion on the case of a scalar resonance, which we will, from here on, denote as $\Phi$.  We will also see that $\Phi$ can be produced from a parent messenger resonance, which we will denote as $M$, and this parent messenger resonance can be spin 1, 0 or 1/2. 

Under the hypothesis that the excess is coming from physics beyond the Standard Model (SM), one should also confront the many other searches for resonant production of a pair of SM particles which constrain possible other decay modes of $\Phi$.
The full list of relevant limits has been collected in Appendix~\ref{app:limits} which provides the description of the inputs in our numerical analysis. Here we summarize only the most important ones, rescaled to 13~TeV rates to facilitate the discussion:

\begin{center}
\begin{tabular}{c|c|c}
Final State & 95\% CL U.L. on $\sigma \times \textrm{BR}$ [fb] & lim. normalized to $\sigma_{\ga\ga} = 5\div10\,{\rm fb}$ \\
\hline
WW (gluon fusion) & 174 & $17.4\div34.8$\\ 
WW (VBF) & 70 & $7\div14$ \\
ZZ (gg prod.) & 89 & $9\div18$ \\
ZZ (VBF prod.) & 40 & $4\div8$ \\
Z$\ga$ & 42 & $4.2\div8.4$ \\
Zh & 572 & $57\div114$\\
hh & 209 & $21\div 42$ \\
bb & $10^{4}$ & $1\div2\times10^{3}$\\
tt & $4.04 \times 10^{3}$ & $404\div807$\\
$\tau\tau$ (gg prod.) & 56 & $6\div11$\\
$\tau\tau$ (assoc. b production) & 54 & $5.4\div10.8$ \\
qq & $10^{4}$ & $1\div2\times10^{3}$\\
$\ell\ell$ & 3.5 & $0.35\div0.7$
\end{tabular}
\end{center}
For convenience we have added a column normalizing the limits to the cross section required by the $\ga\ga$ excess.

From this table it is easy to answer the question of whether it is possible to accommodate the di-photon excess by extending the SM with only one particle. In particular one can see that the above numbers imply the following lower bounds on the di-photon branching ratios, such as
\begin{align}\label{eq:bound-ratio}
\begin{array}{rl}
BR(\Phi\rightarrow \ga\ga)/BR(\Phi\rightarrow W^{+} W^{-}) &\gtrsim 5\div10/174 \sim 2.9\div5.7 \times 10^{-2}\\ BR(\Phi\rightarrow \ga\ga)/BR(\Phi\rightarrow t \bar t) &\gtrsim 5\div10/4036 \sim 1.24\div2.48 \times 10^{-3}
\end{array}
\end{align}
and so on. If a coupling of $\Phi$ to the $t$ and/or $W$ is responsible for a loop induced decay to $\gamma\gamma$, then there is no obstruction for the tree-level decay modes in the denominator of (\ref{eq:bound-ratio}). From simple dimensional analysis, we already see that 
\begin{align}
BR(\Phi\rightarrow \ga\ga)/BR(\Phi\rightarrow W^{+} W^{-} / t \bar t)) \sim \left(\frac{\alpha}{4\pi}\right)^2 \sim5\times 10^{-5}
\end{align}
which is several orders of magnitude smaller than the lower bounds in (\ref{eq:bound-ratio}). For the remaining SM fermions this tension is even stronger due to the chiral suppression in the loop function. In particular for the bottom quark this suppression is more than enough to rule out a bottom-loop induced decay, even though the constraint on $b\bar b$ is somewhat weaker than the constraints on $t\bar t$ and $WW$.
Using the table above one can see that it is not possible to significantly increase the di-photon branching ratio without violating the limits on one of the tree-level decays first.
Thus, we find that decay of the resonance through SM particles is not viable: we need additional new physics! 

\begin{table}
\renewcommand\arraystretch{5}
\begin{tabular}{|>{\centering}m{2in} |>{\centering}m{2in} |>{\centering\arraybackslash}m{1in}|}\hline
\begin{tikzpicture}

\begin{scope}[thick,decoration={
    markings,
    mark=at position 0.8 with {\arrow{stealth}}}
    ]

\draw[-,dashed] (1.5,0.5)--(2.5,0.5); 
 \draw[-] (2.5,0.5)--(3,1); 
 \draw[-] (2.5,0.5)--(3,0); 
 \draw[-] (3,1)--(3,0); 

\draw[photon] (3,1)--(4,1); 
\draw[photon] (3,0)--(4,0); 

\node[black] at (1.3,0.5) {$\Phi$};

\node[black] at (2.5,-0.4) {SM};

 \draw[->] (2.5,-0.2)--(2.7,0.2);

\end{scope}

\end{tikzpicture}&The SM + $\Phi$ is not viable \newline& Tree-level decays of SM particles excluded by LHC searches \newline\\\hline
\begin{tikzpicture}
[scale=0.75]
\begin{scope}[thick,decoration={
    markings,
    mark=at position 0.8 with {\arrow{stealth}}}
    ] 
    
\draw[gluon] (0,1)--(1,1); 
\draw[gluon] (0,0)--(1,0); 

\draw[-] (1,0)--(1,1); 
\draw[-] (1,0)--(1.5,0.5); 
\draw[-] (1,1)--(1.5,0.5); 
\draw[-,dashed] (1.5,0.5)--(2.5,0.5); 
 \draw[-,very thick, red,double=black,double distance=1pt] (2.5,0.5)--(3,1); 
 \draw[-,very thick, red,double=black,double distance=1pt] (2.5,0.5)--(3,0); 
 \draw[-,very thick, red,double=black,double distance=1pt] (3,1)--(3,0); 
\draw[photon] (3,1)--(4,1); 
\draw[photon] (3,0)--(4,0); 

 \node[black] at (1.4,-0.) {$t$};
 \node[black] at (3.0,-0.9) {NP};
 \node[black] at (2,0.8) {$\Phi$};
 \node[black] at (0.6,-2.25) {$W,Z$};
 \node[black] at (2.35,-1.95) {$\Phi$};

 \draw[->] (2.8,-0.5)--(2.7,-0.); 
 \draw[->] (3,-1.2)--(3.,-1.7);

\draw[-] (0,-1.5)--(1,-1.8); 
\draw[-] (0,-3)--(1,-2.8); 
\draw[-] (1,-1.8)--(2,-1.5); 
\draw[-] (1,-2.8)--(2,-3); 

\draw[photon] (1,-2.8)--(2,-2.25); 
\draw[photon] (1,-1.8)--(2,-2.25); 
\draw[-,dashed] (2,-2.25)--(2.75,-2.25);

 \draw[-,very thick, red,double=black,double distance=1pt] (2.5+0.25,0.5-2.75)--(3+0.25,1-2.75); 
 \draw[-,very thick, red,double=black,double distance=1pt] (2.5+0.25,0.5-2.75)--(3+0.25,0-2.75); 
 \draw[-,very thick, red,double=black,double distance=1pt] (3+0.25,1-2.75)--(3+0.25,0-2.75); 
\draw[photon] (3+0.25,1-2.75)--(4+0.25,1-2.75); 
\draw[photon] (3+0.25,0-2.75)--(4+0.25,0-2.75);

\end{scope}

\end{tikzpicture}&SM ggF$\to \Phi$ and VBF$\to \Phi$\newline& Rate to di-photons requires large 't Hooft coupling, eq.~\eqref{constraint} \newline\\\hline
\end{tabular}
\caption{The SM + $\Phi$ is not viable. SM gluon fusion and VBF  production requires boosting the decay width to di-photons via a large 't Hooft coupling. \label{tab:SMplusX}}
\renewcommand\arraystretch{1.2}
\end{table}

The next consideration is whether production can be SM-like, while the decay to $\gamma \gamma$ occurs through loops of heavy messengers.  To address this question, we introduce the rate of resonant $\Phi$ production and decay to di-photons:
\begin{align}
R_{ \gamma\gamma}=&\,\hat\sigma_{in \rightarrow \Phi, incl.} m_{\Phi}^2 \,\frac{\Gamma_{ \gamma\gamma}}{\Gamma_{ \gamma\gamma}+\Gamma_{other}}\,\frac{d\mathcal{L}}{d m_{\Phi}^2} \,,
\label{eq:four}
\end{align}
where $\hat\sigma_{in \rightarrow \Phi, incl}$ is the inclusive, parton-level cross section for a particular initial state `$in$'\footnote{In this paper, we use leading order estimates, and do not include $K$ factors for the production cross-sections. The effect of $K$ factors will increase the quoted rates by factors of up to two.}, $\Gamma_{\gamma\gamma}$ is the partial width to $\gamma\gamma$, $\Gamma_{other}$ is the total width from any other decays, and the final factor is the relevant parton luminosity function evaluated at the mass of $\Phi$ ($m_{\Phi}$). 
 
 For a two-body initial state, SM SM$\to\Phi$, to leading order we can then always rewrite this in terms of the decay width of the process $\Phi\rightarrow in$: 
\begin{align}\label{eq:rateintermsofwidth}
R_{ \gamma\gamma}\sim &\, \frac{\Gamma_{ in}}{m_{\Phi}} \frac{\Gamma_{ \gamma\gamma}}{\Gamma_{\gamma\gamma}+\Gamma_{other}}\,\frac{d\mathcal{L}}{d m_{\Phi}^2} \,,
\end{align}
where $\sim$ denotes some ${\cal O}(1)$, process dependent, symmetry factors. Eq.~(\ref{eq:rateintermsofwidth}) is crucial, as it connects the rate of production with the width of the resonance. The width/rate interplay is of great importance, as a decay to di-photon is normally at the loop level.\footnote{Eq.~(\ref{eq:rateintermsofwidth}) only assumes resonant production in the narrow width approximation (NWA), which is supported by the data that prefers at most $\Gamma/M \sim 6\%$}. Let us see how this matters in addressing the possibility of SM-like production. Since $\Phi$ can always decay back to the initial state, we have
\begin{equation}\label{inequality}
\Gamma_{other}\geq \Gamma_{in}>0.
\end{equation}
Let us now consider for instance production through gluon fusion by coupling the $\Phi$ to the SM top. In this case we have $\Gamma_{other}\approx \Gamma_{tt}\gg \Gamma_{in}=\Gamma_{gg}$. The expression then becomes
\begin{align}
R_{ \gamma\gamma}\sim &\,  \frac{\Gamma_{gg}}{\Gamma_{tt}}\frac{\Gamma_{\gamma\gamma}}{m_{\Phi}}\,\frac{d\mathcal{L}}{d m_{\Phi}^2} \,.
\label{constraint}
\end{align}
With $\Gamma_{gg}/\Gamma_{tt}\sim 10^{-3}$, and the parton luminosity for a resonance of mass $M_{\Phi}=750\,{\rm GeV}$
\begin{align}
R_{ \gamma\gamma}\sim &\, 10^{-3}\times \frac{\Gamma_{\gam\gam}}{750\, \text{GeV}}\times10^6\, \text{fb}\sim \,\frac{\Gamma_{\gam\gam}}{ \text{GeV}} \,\text{fb}.
\end{align}
That is, to obtain the observed rate we need a  partial width to $\gamma\gamma$ of order 1\,GeV. As we will see in the next section, the typical partial width to $\gamma\gamma$ from a loop of messengers is $\sim 1$ MeV or smaller, and so this width needs to be boosted in some way. Adding a large number  of messengers pushes the theory to the strongly coupled regime. Given that ATLAS data slightly prefers a largish width, ${\cal O}(6\%)$, discussing possible avenues to achieve it, is of some importance, especially given the model building challenges. Therefore  the total width will be the subject of the following section.

Another SM-only production possibility is vector boson fusion. In this case the production is suppressed by the three-body phase space rather than by a loop factor. Following a similar argument to the above, it is easy to see that a partial width  to $\gamma\gamma$ of order 1\,GeV is again required.

\renewcommand\arraystretch{5}
\begin{table}[h]
\begin{tabular}{|>{\centering}m{2in} |>{\centering}m{2in} |>{\centering\arraybackslash}m{1in}|}\hline
\begin{tikzpicture}

\begin{scope}[thick,decoration={
    markings,
    mark=at position 0.8 with {\arrow{stealth}}}
    ] 
    
\draw[gluon] (0,1)--(1,1); 
\draw[gluon] (0,0)--(1,0); 

\draw[-,very thick, red,double=black,double distance=1pt] (1,0)--(1,1); 
\draw[-,very thick, red,double=black,double distance=1pt] (1,0)--(1.5,0.5); 
\draw[-,very thick, red,double=black,double distance=1pt] (1,1)--(1.5,0.5); 
\draw[-,dashed] (1.5,0.5)--(2.5,0.5); 
 \draw[-,very thick, red,double=black,double distance=1pt] (2.5,0.5)--(3,1); 
 \draw[-,very thick, red,double=black,double distance=1pt] (2.5,0.5)--(3,0); 
 \draw[-,very thick, red,double=black,double distance=1pt] (3,1)--(3,0); 

\draw[photon] (3,1)--(4,1); 
\draw[photon] (3,0)--(4,0); 

 \node[black] at (2.0,-0.3) {NP};
 \node[black] at (2,0.75) {$\Phi$};

\draw[->] (2.2,-0.1)--(2.6,0.2); 
\draw[->] (1.8,-0.1)--(1.4,0.2);

\end{scope}

\end{tikzpicture}&Gluon fusion through a heavy colored messenger\newline&Section \ref{sec:coloredmess} \newline\\\hline
\begin{tikzpicture}
[scale=1.3]
\begin{scope}[thick,decoration={
    markings,
    mark=at position 0.8 with {\arrow{stealth}}}
    ]

\draw[-] (0,1)--(0.5,0.5); 
\draw[-] (0,0)--(0.5,0.5); 
\draw[-,dashed] (0.5,0.5)--(1.5,0.5); 

 \draw[-,very thick, red,double=black,double distance=1pt] (1.5,0.5)--(2,1); 
 \draw[-,very thick, red,double=black,double distance=1pt] (1.5,0.5)--(2,0); 
 \draw[-,very thick, red,double=black,double distance=1pt] (2,1)--(2,0); 
\draw[photon] (2,1)--(3,1); 
\draw[photon] (2,0)--(3,0); 

 \node[black] at (1.2,-0.2) {NP};
 \node[black] at (0.1,-0.1) {$q$};
 \node[black] at (1,0.7) {$\Phi$};
 
\draw[->] (1.4,-0.05)--(1.65,0.2);

\end{scope}

\end{tikzpicture}&Non-MVF Yukawa coupling to first generation quarks\newline&Section \ref{subsec:uncolored}\newline\\\hline
\begin{tikzpicture}
[scale=1.3]
\begin{scope}[thick,decoration={
    markings,
    mark=at position 0.8 with {\arrow{stealth}}}
    ]

\draw[-] (0,1)--(0.5,0.5); 
\draw[-] (0,0)--(0.5,0.5); 
\draw[-] (0.5,0.5)--(1,1); 
\draw[-] (0.5,0.5)--(1,0); 
\draw[-,dashed] (0.5,0.5)--(1.3,0.5); 

 \node[black] at (0.1,-0.1) {$q$};
 \node[black] at (1.3,0.7) {$\Phi$};

\end{scope}

\end{tikzpicture}&Vector boson fusion through a heavy $W'$\newline&Appendix \ref{sec:VBF}\newline\\\hline\hline
\begin{tikzpicture}

\begin{scope}[thick,decoration={
    markings,
    mark=at position 0.8 with {\arrow{stealth}}}
    ]

\draw[-,very thick, red,double=black,double distance=1pt] (1.5,0.5)--(2.5,0.5); 
\draw[-] (2.5,0.5)--(3.,1); 
\draw[-,dashed] (2.5,0.5)--(3.,0); 

\draw[photonvalley] (3,0)--(3.7,-0.55); 
\draw[photonvalley] (3,0)--(2.9,-0.8); 
    
    \node[black] at (2.6,0.1) {$\Phi$};
    \node[black] at (2.6,1.1) {SM};

        \node[black] at (1.5,0.1) {NP};

\end{scope}

\end{tikzpicture}&Cascade decay\newline&Section \ref{sec:cascades}\newline\\\hline
\begin{tikzpicture}

\begin{scope}[thick,decoration={
    markings,
    mark=at position 0.8 with {\arrow{stealth}}}
    ]

\draw[-,very thick, red,double=black,double distance=1pt] (1.5,0.5)--(2.5,0.5); 
\draw[photonvalley] (2.5,0.5)--(3.,1); 
\draw[-,very thick, red,double=black,double distance=1pt] (2.5,0.5)--(3.,0); 

\draw[photonvalley] (3,0)--(3.7,-0.55); 
\draw[-] (3,0)--(2.9,-0.8); 
    
    \node[black] at (2.4,0.1) {NP};

        \node[black] at (1.5,0.1) {NP};
        \node[black] at (2.6,-0.8) {SM};

\end{scope}

\end{tikzpicture}&Non-resonant kinematic edge providing excess\newline&Section \ref{sec:cascades}\newline\\\hline
\begin{tikzpicture}

\begin{scope}[thick,decoration={
    markings,
    mark=at position 0.8 with {\arrow{stealth}}}
    ] 
    
%
\draw[-,dashed] (1.5,0.5)--(2.5,0.5); 
\draw[-,dashed] (2.5,0.5)--(3.,1); 
\draw[-,dashed] (2.5,0.5)--(3.,0);

\draw[photonvalley] (3,1)--(3.4,1.55); 
\draw[photonvalley] (3,1)--(3.6,1.4); 
\draw[photonvalley] (3,0)--(3.4,-0.55); 
\draw[photonvalley] (3,0)--(3.6,-0.4);

\node[black] at (1.3,0.7) {$\Phi$};

\node[black] at (2.65,0.9) {$a$};
\node[black] at (2.65,0.1) {$a$};

\end{scope}

\end{tikzpicture}&Decay to two pairs of collimated photons through a Hidden Valley\newline&Section \ref{sec:HV}\newline\\\hline
\end{tabular}
\caption{Topologies considered in this paper.  \label{tab:topologies}}
\end{table}
\renewcommand\arraystretch{1.2}

We consider five ways forward to generate the observed rate:
\begin{itemize}
\item We can approximately saturate the first inequality in (\ref{inequality}), by ensuring that there are no other important modes for $\Phi$ to decay to other than $\gamma\gamma$ and back to the initial state. In this case \emph{the dependence on the production mechanism cancels from the rate:}
\begin{align}
R_{ \gamma\gamma}\sim &\, \frac{\Gamma_{\gamma\gamma}}{m_{\Phi}}\,\frac{d\mathcal{L}}{d m_{\Phi}^2}\sim  \frac{10^{-3}\text{\,GeV}}{750\, \text{GeV}}\times 10^6\, \text{fb}\sim \,1\, \text{fb}
\end{align}
which is in the right ballpark to explain the excess. In Sec.~\ref{sec:heavy} we present two examples of this kind: in the first example $\Phi$ is produced through gluon fusion induced by a heavy messenger. In the second case the production occurs through a Yukawa coupling to the first generation quarks and the decay through an uncolored messenger.
\item One could accept the suppression inherent to SM-like gluon or vector-boson fusion, and instead attempt to increase $\Gamma_{\gamma\gamma}$ several orders of magnitude. In terms of a loop with messengers, this corresponds to the limit of large `t Hooft coupling and the theory becomes strongly coupled.  This case may be of interest for composite models. Alternatively, in Hidden Valley (HV) models \cite{Strassler:2006ri,Strassler:2006im}, $\Phi$ may decay to very light states, which then each can decay to two very collimated photons, resolved only as a single photon. In this case the analogue of $\Gamma_{\gamma\gamma}$ corresponds to a tree-level decay, and can therefore be much larger. We discuss this case in Sec.~\ref{sec:HV}.
\item The analysis above does not apply for cascade decays, in which $\Phi$ may be produced as the daughter of some heavier messenger resonance, $M$. In this case it is straightforward to increase the production rate of $\Phi$ without decreasing its branching ratio to $\gamma\gamma$. This will however naturally lead to a signature different from a di-photon resonance alone, {\it e.g.} extra jets, ($t$,$b$), MET, leptons. We may infer from the lack of such information in the public results that no such signature is present at a striking level,\footnote{This information has been explicitly confirmed in by the speakers in the public talk at the CERN December 15th, 2015 event.} and therefore that the separation between $\Phi$ and $M$ should not be large and that the extra activity produced in $M$ decays should be predominantly hadronic. We will nevertheless consider this scenario in its full generality, as it provides a natural explanation, and as the additional signatures may not be apparent with any certainty due to small statistics/squeezed spectra.
\item Another variation of the cascade decays is given by the possibility that the di-photon ``peak'' may be a kinematic edge, hard to distinguish due to the relatively low statistics. This provide a natural explanation for the peak ``width'' and the production rate can be easily controlled because it can proceed at tree level.
\item Finally, we consider vector-boson fusion induced by a set of new vector bosons, which are too heavy to contribute at tree-level to the width of $\Phi$. This scenario is, however, already excluded by existing di-jet constraints, and we relegate it to Appendix~\ref{sec:VBF}.
\end{itemize}
The various topologies we consider in this paper are summarized in Table \ref{tab:topologies}. 

A key result of the observation of a di-photon excess is that in all cases we can think of, is we need more new physics beyond the single resonance.  We now turn to discussing the width of the excess, which has important consequences.

\section{Importance of the Width}
\label{sec:width}

Early indications, driven by ATLAS, are that the new resonance may have a substantial width, ${\cal O}(6\%)$.  Since the decay to $\ga\ga$ is a loop process and is naturally small, the observation of a substantial width has important implications for the theory.  We discuss these separately for the $pp \rightarrow \Phi \rightarrow \ga\ga$ case (explored in more detail in Sec.~\ref{sec:heavy}) and for the cascade decay case (discussed in more detail in Sec.~\ref{sec:cascades}; the conclusions on the width for the cascade case will also apply to the Hidden Valley of Sec.~\ref{sec:HV}).   

\subsection{$pp \rightarrow \Phi \rightarrow \ga\ga$ process}

As we have seen, the rate in the $pp \rightarrow \Phi \rightarrow \ga\ga$ process is given by 
\begin{equation}
R_{ \gamma\gamma}\sim \, \frac{1}{m_{\Phi}} \frac{\Gamma_{in}\Gamma_{ \gamma\gamma}}{\Gamma_{\gamma\gamma}+\Gamma_{in}+\delta\Gamma}\,\frac{d\mathcal{L}}{d m_{\Phi}^2} = 5\div10 \mbox{ fb}
\end{equation}
where $\delta\Gamma$ is the partial width into states not involved in production or $\ga\ga$ decay. 

\begin{figure}
\includegraphics[width=0.45\textwidth]{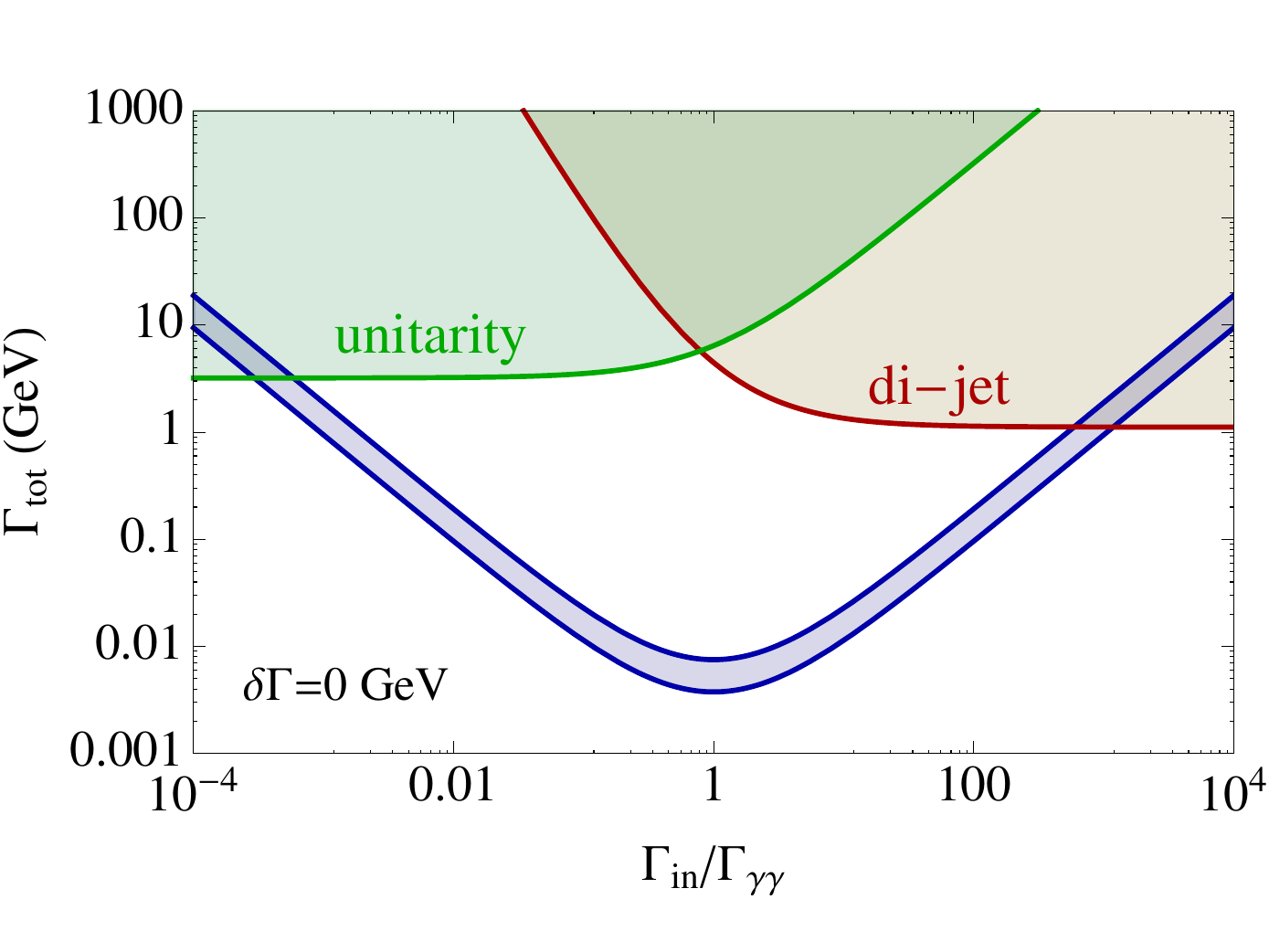}\qquad
\includegraphics[width=0.45\textwidth]{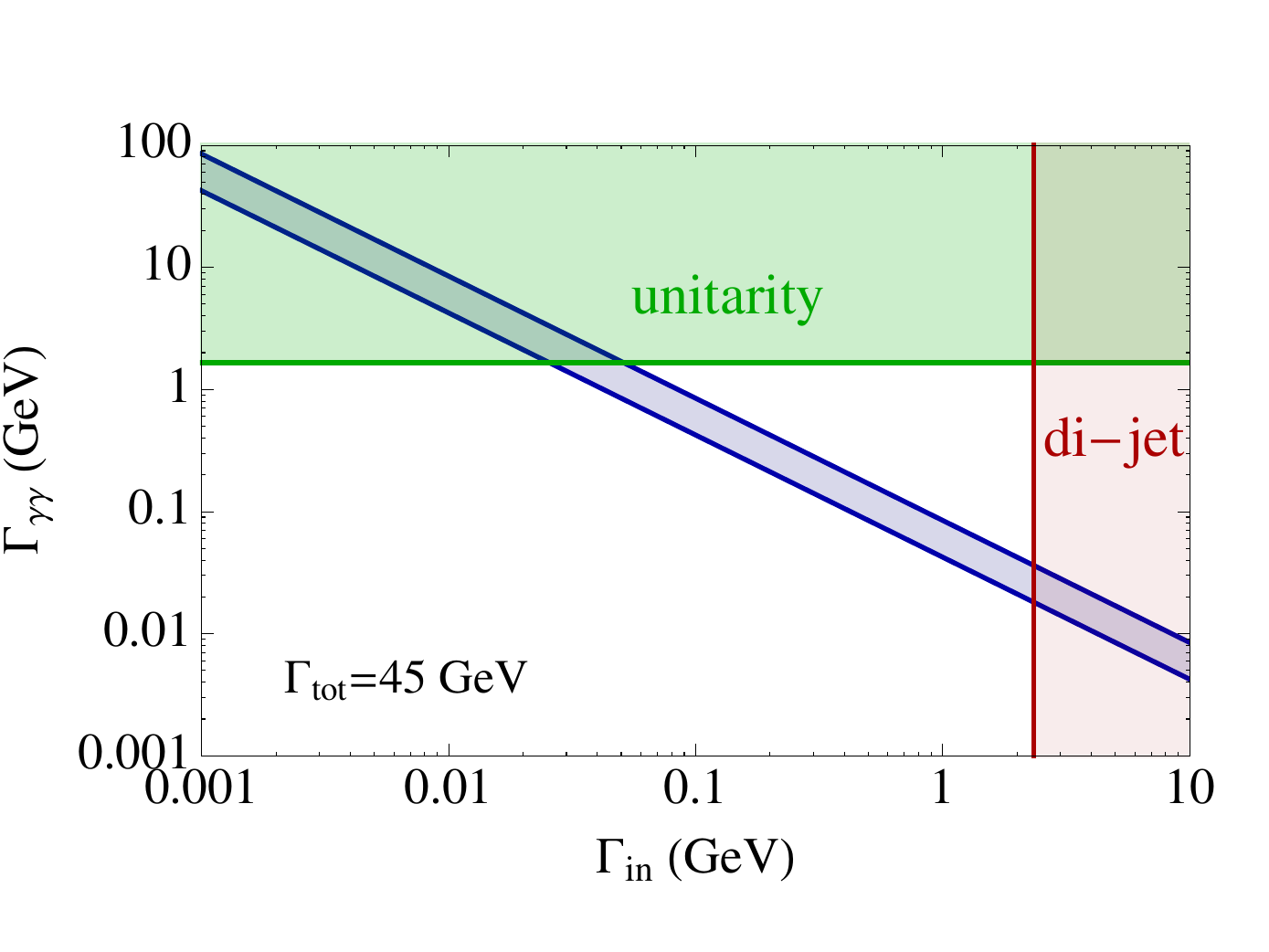}
\caption{{\em Left:} Allowed width ranges for explaining the di-photon resonance, in blue, assuming the production and decay dominate the total width of the resonance.  Constraints from unitarity (for reference model $F9$ of Table~\ref{SingletBranchingRatios}) and di-jet are shown as shaded regions. {\em Right:} Same as left panel, but fixing the total width to 45~GeV by allowing for other, unobserved decay modes. In both plots a $q\bar q$ initial state was assumed; the results for a $gg$ initial state are qualitatively similar.\label{moneyplot}}
\end{figure}

If we hold $R_{\gamma\gamma}$ fixed to fit the excess, we can solve for $\Gamma_{\ga\ga}$ as a function of $\Gamma_{in}$ and vice versa.  This is shown in Fig.~\ref{moneyplot}, as a blue band.  (In this figure we assumed a $q\bar q$ initial state, as this provides somewhat more freedom in terms of varying $\Gamma_{in}$, see section \ref{subsec:uncolored}.) Consider first the left-hand panel, in which $\delta\Gamma = 0$.  If we increase $\Gamma_{in}$, it drops out from the expression, and the branching ratio to $\ga\ga$ is very small, but compensated by the large production rate. The total width of the resonance also grows, as it is dominated by $\Gamma_{in}$. Eventually this direction is cut off by the constraints on di-jet resonances (red region in Fig.~\ref{moneyplot}). Similarly, if we increase $\Gamma_{\ga\ga}$, we eventually approach the point where nearly 100\% branching ratio is to $\ga\ga$. This direction is, however, bounded by unitarity considerations, since at some point the `t Hooft coupling becomes non-perturbative and the theory enters a regime of strong dynamics. This constraint is of course model dependent, and is shown in the green region in Fig.~\ref{moneyplot} for one of the models studied in Sect.~\ref{subsec:uncolored}, $F9$ of Table~\ref{SingletBranchingRatios}.  Notice that the left-hand panel implies that it is hard to obtain a 45~GeV width for the particle when the only contribution to its width is through the production and decay channels.  

Next, consider the impact of adding a decay of $\Phi$ to states not initiating the production or decay.  This is shown in the right-hand panel of Fig.~\ref{moneyplot}, where $\Gamma_{tot} = \Gamma_{\gamma\gamma}+\Gamma_{in}+\delta\Gamma = 45$~GeV.  In this case, either one requires the width to $\ga\ga$ to be very large ($\gg$ 1 MeV, which is an upper bound on the natural decay width through a single charged loop of fermions), or $\Gamma_{in}$ must be substantial itself.  However even in the latter case, Fig.~\ref{moneyplot} indicates that $\Gamma_{in}$ is bounded by di-jet constraints, and we always have $\Gamma_{in}\ll45$~GeV. This implies that the total width is always dominated by the exotic decay modes, parametrized by $\delta \Gamma$.

We conclude that there are two possibilities: {\em (i)} the particle is narrow and its width can be dominated by the decay to $\gamma\gamma$, the decay back the initial state or a combination of both, or {\em (ii)} the particle is broad in which case a substantial range of partial widths to the initial state and to $\ga\ga$ are possible.   However a total width of 45~GeV cannot be obtained from $\Gamma_{in}+\Gamma_{\ga\ga}$, whether that be at loop or tree level, due to unitarity and di-jet constraints. A sizable partial width to other states is therefore needed.  In Sec.~\ref{sec:heavy}, we explore what values for $\Gamma_{in},\Gamma_{\ga\ga}$ are possible in concrete models.

\subsection{$pp \rightarrow M \rightarrow \Phi(\rightarrow \ga\ga) + SM$ process}

\begin{figure}
\includegraphics[width=0.45\textwidth]{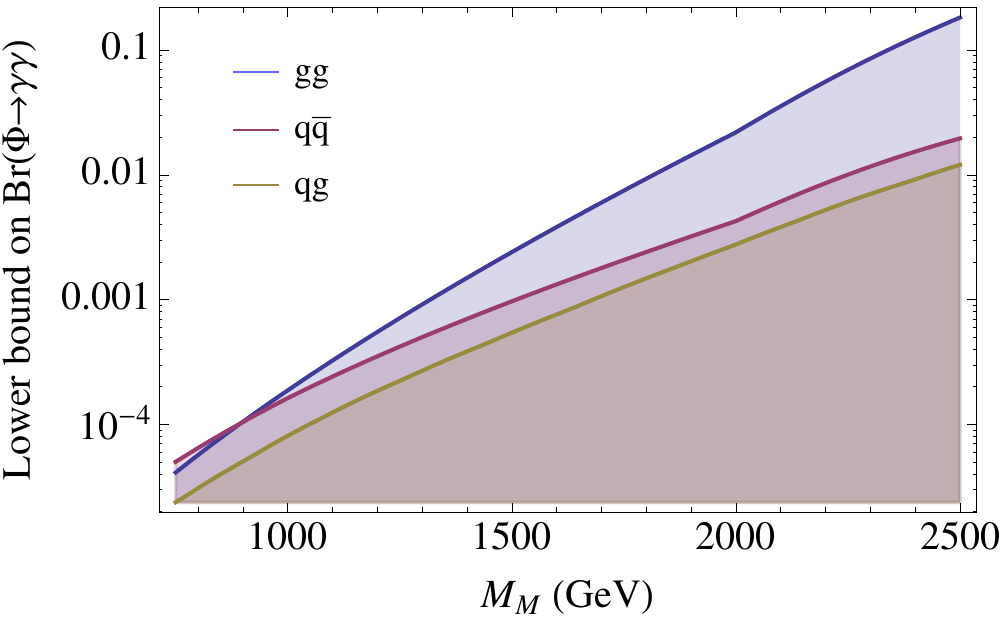}
\caption{Lower bound on the branching fraction of $\Phi \rightarrow \ga\ga$, for various parton luminosities in the initial state, as a function of the messenger mediator mass, $m_M$, which decays $M \rightarrow \Phi + \mbox{ SM}$, with $\Phi \rightarrow \ga\ga$. We fix the ratio $\Gamma^M_{tot}/m_M=0.1$. \label{moneyplotCascade}}
\end{figure}

Next we consider the importance of the width to $\ga\ga$ in the case that a messenger resonance, $M$, is produced first, which decays to $\Phi$ plus the SM.  Here one is able to better factorize production from the branching fraction of $\Phi$, though even in this case we will find constraints on the branching ratio of $\Phi \rightarrow \ga\ga$.  First, again using the narrow width approximation, Eq.~(\ref{eq:rateintermsofwidth}), and expressing the production cross section in terms of the partial width of the messenger $M$ to decay to jets, we can express the total di-photon rate $R_{\ga\ga}$ in terms of a product of branching ratios. We therefore require
\be
\frac{\Gamma^{M}_{tot}}{m_M}\left(\frac{d\mathcal{L}}{d m_M^2}  c \right) BR(M\rightarrow j j) BR(M\rightarrow \Phi +\mbox{SM})BR(\Phi\rightarrow \ga\ga) =5\div10\,{\rm fb}
\label{BrBrBr}
\ee
where we have converted $\Gamma(M\rightarrow j j)$ into a branching ratio and made the total width of the $M$ explicit.
The constant $c$ is a numerical factor that for a $q\bar q$ initial state is $c_{q\bar q} = 4\pi^{2}/9$. In the following we will also use $c_{q g} = \pi^{2}/6$ and $c_{g g} = \pi^{2}/8$ for the quark-gluon and gluon-gluon initial states respectively.  Next, we can maximize the messenger production rate times branching fraction to $\Phi$ by choosing $BR(M\rightarrow j j) = BR(M\rightarrow \Phi +\mbox{SM}) = \frac{1}{2}$.  This then allows us to put a lower bound on the branching fraction $BR(\Phi\rightarrow \ga\ga)$, as a function of $m_M$, for various initial state parton luminosities.  This is shown in Fig.~\ref{moneyplotCascade}.

From this plot, we immediately see that if $\Phi$ has a large width (45~GeV), the absolute width to $\ga\ga$ is bounded to be 1 MeV or larger for $m_M = 800$~GeV, but rapidly increases with $M$.  For example, adding a single charged fermion with a coupling $\sim 1$ to $\Phi$ naturally generates a width to $\ga\ga$ of at most 1 MeV.  Increasing the mass of the messenger necessitates exponentially larger values of $\Gamma_{\ga\ga}$, rendering the structure of the model progressively more complicated. Therefore we can see that the sizable width of $\Phi$ prefers lighter messenger masses, which in turn is consistent with the absence of extra energetic objects in the events. Nevertheless, even in this case, the dominant width of $\Phi$ must be to an exotic or hidden channel.  The reason for this is that searches for resonances in other SM channels may strongly constrain $BR(\Phi \rightarrow \ga\ga)$.  For example, the di-jet constraint implies that $BR(\Phi \rightarrow \ga\ga)/BR(\Phi \rightarrow jj) \gtrsim 10^{-3}$, so if the dominant decay is to SM states, $\Gamma_{\gamma\gamma} \gtrsim 10^{-3} \Gamma_{tot}$.  In this case, many charged particles (and/or particles with large electric charge) must be present to boost the width of $\Gamma_{\gamma\gamma}$ to 10's of MeV.

\section{Heavy Messengers}
\label{sec:heavy}

We established in the introduction that a viable model of $\Phi \rightarrow \gamma \gamma$ requires new states in addition to the resonance itself.  
In this section we present a class of models which explicitly realize this scenario as a two-to-two $pp \rightarrow \Phi \rightarrow \ga\ga$. (This scenario was examined in detail prior to the appearance of the di-photon excess in terms of effective field theory \cite{Jaeckel:2012yz}. In this paper we pursue an analysis in terms of simplified models, since in the case at hand it is particularly straightforward to get a complete picture by interpolating between a sufficiently complete set of such simplified models.)

Messenger multiplets are highly motivated in a wide variety of physics beyond the standard model, from Grand Unified Theories (GUTs) to composite sectors.  
A simple extension, that encompasses many models, is an additional vector-like fermion\footnote{See \cite{Aguilar-Saavedra:2013qpa} for an earlier study of new vector-like quark multiplets.},
\begin{equation}\label{lagferm}
\mathcal{L}= \mathcal{L}_{SM}+\frac{1}{2}m_\Phi^2\Phi^2 +(g_f\Phi+m) \bar\Psi\Psi 
\end{equation}
or a complex scalar,
\begin{equation}\label{lagscal}
\mathcal{L}=\mathcal{L}_{SM}+\frac{1}{2}m_\Phi^2\Phi^2 +(g_s \Phi+m^2) |\phi|^2. 
\end{equation}
Because the decay to photons is loop suppressed in the class of models we consider here, if the resonance has a substantial width, the phenomenology prefers that there be a high multiplicity of these messenger particles, possibly motivated by compositeness.
In addition, when $m<m_{\Phi}/2$, a tree level decay into $\Psi$ or $\phi$ pairs is opened, diluting the partial width to $\ga\ga$, and leading to an interesting new signature, $\Phi \rightarrow \Psi \bar \Psi$ or $\Phi \rightarrow \phi \phi$.

We first consider the case of a colored messenger multiplet, where the messenger participates in both the production and decay of the new resonance.  Then we consider the case that the production is through a tree level coupling to the SM and the messengers mediate the decay only. 

\subsection{A Colored Messenger Multiplet}
\label{sec:coloredmess}
We first consider the case that $\Psi$ and $\phi$ are color triplets, have a fast, tree-level decay to the SM and allow for $N_f$ flavors for $\Psi$ and $\phi$. With these conditions, we find 10 possible representations for $\phi$ and 7 for $\Psi$, as summarized in Table~\ref{TripletBranchingRatios}.   Each entry of this table can be thought of as a simplified model with parameters $g$, $m$ and $N_f$. As a convention, we only consider `holomorphic' couplings of the messengers with the standard model matter fields. For instance, while S1 and S2 contribute identically to the decay and production of $\Phi$, the decay mode of $\phi$ is given by the operators $\phi d^c e^c$ and $\phi u^c u^c$ respectively. Both the production and the decay of $\Phi$ will then occur through a loop of $\Psi$ or $\phi$ states. In addition to $\gamma\gamma$, other possible decay modes are $gg$, $WW$ and $Z\gamma$, depending on the representations.

We assume that $m > m_\Phi/2$, so that $\Phi$ has no tree-level decay modes, and the $\gamma\gamma$ channel is relatively unsuppressed relative to these other channels. The ratios of the leading order partial widths are shown in Table~\ref{TripletBranchingRatios}. We see that it is usually a good approximation to take $\Gamma_{gg}\gg \Gamma_{\gamma\gamma},\Gamma_{Z\gamma},\Gamma_{WW},\Gamma_{ZZ}$, such that (\ref{eq:rateintermsofwidth}) simplifies to
\begin{align}
R_{ \gamma\gamma}\approx&\frac{\pi^2}{8 m_{\Phi}}\,\Gamma_{ \gamma\gamma}\frac{d\mathcal{L}^{gg}}{d m_{\Phi}^2}
\label{eq:sec2arate}
\end{align}
where we included the appropriate symmetry factors that were omitted in (\ref{eq:rateintermsofwidth}).

\begin{table}[t]
\begin{eqnarray*}
\begin{array}{|c|c|c|c|c|c|c|c|c|c|}
\hline
\text{Model}& \text{Representation} &  \text{$\gamma $Z/$\gamma \gamma $} &
   \text{WW/$\gamma \gamma $} &\text{ZZ/$\gamma \gamma $} &
   \text{gg/$\gamma \gamma $} & R^0_{\Phi\rightarrow\gamma\gamma}\,\text{[fb]} &\Gamma_{tot}\, \text{[MeV]} & \Gamma_{\Phi\rightarrow \gamma\gamma}\,\text{[MeV]} &
 \mbox{Decay mode} \\ \hline \hline
   \multicolumn{10}{|c|}{{\rm Scalars}} \\ \hline \hline
 \text{S1} & \left(3,1,-\frac{4}{3}\right) & 0.6 & 0. & 0.09 & 9.54 & 0.02 &
   0.03 & 3.\times 10^{-3} & d^c\text{ + }e^c \\
 \text{S2} & \left(\bar{3},1,\frac{4}{3}\right) & 0.6 & 0. & 0.09 & 9.54 & 0.02 &
   0.03 & 3.\times 10^{-3} & \text{2 }u^c \\
 \text{S3} &  \left(3,2,\frac{7}{6}\right)& 0.06 & 0.91 & 0.6 & 11.62 & 0.06 &
   0.14 & 9.9\times 10^{-3} & u^c\text{ + l} \\
 \text{S4} & \left(\bar{3},2,-\frac{7}{6}\right)  & 0.06 & 0.91 & 0.6 & 11.62 & 0.06
   & 0.14 & 9.9\times 10^{-3} & e^c\text{ + q} \\
 \text{S5} &\left(\bar{3},3,\frac{1}{3}\right)& 4.44 & 27.78 & 8.48 & 49.84 & 0.02
   & 0.47 & 5.2\times 10^{-3} & \text{q + l} \\
 \text{S6} &   \left(3,3,-\frac{1}{3}\right)  & 4.44 & 27.78 & 8.48 & 49.84 &
   0.02 & 0.47 & 5.2\times 10^{-3} & \text{2 q} \\
 \text{S7} &  \left(\bar{3},1,-\frac{2}{3}\right) & 0.6 & 0. & 0.09 & 1.5\times 10^2
   & 1.4\times 10^{-3} & 0.03 & 1.9\times 10^{-4} & \text{2 }d^c \\
 \text{S8} & \left(3,2,\frac{1}{6}\right) & 5.07 & 30.62 & 9.26 & 3.9\times
   10^2 & 2.\times 10^{-3} & 0.13 & 2.9\times 10^{-4} & d^c\text{ + l} \\
 \text{S9} &\left(3,1,-\frac{1}{3}\right) & 0.6 & 0. & 0.09 & 2.4\times 10^3
   & 8.7\times 10^{-5} & 0.03 & 1.2\times 10^{-5} & e^c\text{ + }u^c \\
 \text{S10} & \left(\bar{3},1,\frac{1}{3}\right) & 0.6 & 0. & 0.09 & 2.4\times 10^3
   & 8.7\times 10^{-5} & 0.03 & 1.2\times 10^{-5} & d^c\text{ + }u^c \\
   \hline\hline
    \multicolumn{10}{|c|}{{\rm Fermions}} \\ \hline
   \hline
 \text{F1} &  \left(3,2,\frac{7}{6}\right)  & 0.06 & 0.91 & 0.6 & 11.62 & 3.52 &
   8.19 & 0.58 & u^c\text{ + V/h} \\
 \text{F2} &\left(\bar{3},3,-\frac{2}{3}\right) & 1.55 & 13.61 & 4.53 & 24.42  &
   2.49 & 27.86 & 0.62 & \text{q + V/h} \\
 \text{F3} & \left(3,2,-\frac{5}{6}\right) & 0.01 & 2.65 & 1.22 & 33.8 &  1.29
   & 7.67 & 0.2 & d^c\text{ + V/h} \\
 \text{F4} &\left(\bar{3},3,\frac{1}{3}\right)& 4.44 & 27.78 & 8.48  & 49.84 & 1.23
   & 27.7 & 0.3 & \text{q + V/h} \\
 \text{F5} & \left(\bar{3},1,-\frac{2}{3}\right) & 0.6 & 0.  & 0.09 & 1.5\times 10^2
   & 0.08 & 1.69 & 0.01 & \text{q + V/h} \\
 \text{F6} & \left(3,2,\frac{1}{6}\right)& 5.07 & 30.62  &
   9.26 & 3.9\times 10^2 & 0.11 & 7.49 & 0.02 & u^c\text{ + V/h} \\
 \text{F7} &\left(\bar{3},1,\frac{1}{3}\right) & 0.6 & 0.  & 0.09 & 2.4\times 10^3&
   5.1\times 10^{-3} & 1.68 & 6.9\times 10^{-4} & \text{q + V/h} 
 \\ \hline
\end{array}
\end{eqnarray*}
\caption{Quantum numbers of the models we consider and their leading order branching fractions for various final states of the $\Phi$ decay.  The upper part of the table is for scalar loops ($\phi$), while the lower part is for fermion loops ($\Psi$). We include the di-photon rate ($R^0_{\Phi\rightarrow\gamma\gamma}$, in fb), the total width ($\Gamma_{tot}$, in MeV) and width to photons ($\Gamma_{\gam\gam}$, in MeV) for a benchmark point with $m=g_s=1$ TeV, $g_f=1$ and $N_f=1$. Shown alongside the branching ratios are the decay modes for $\Psi$/$\phi$, where $V$ stands for $W$ or $Z$. }
\label{TripletBranchingRatios}
\end{table}

\begin{figure}
\includegraphics[width=0.4\textwidth]{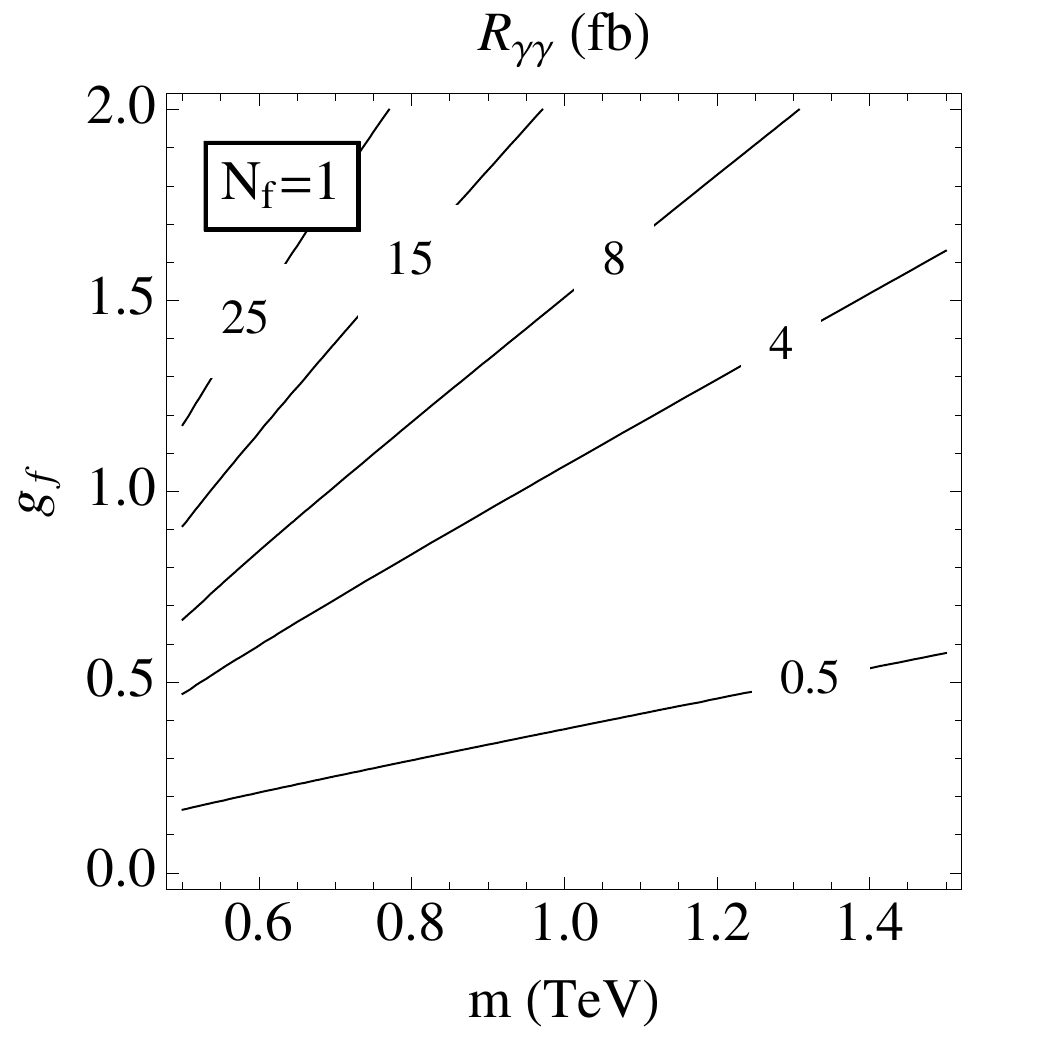}\hspace{1cm}
\includegraphics[width=0.4\textwidth]{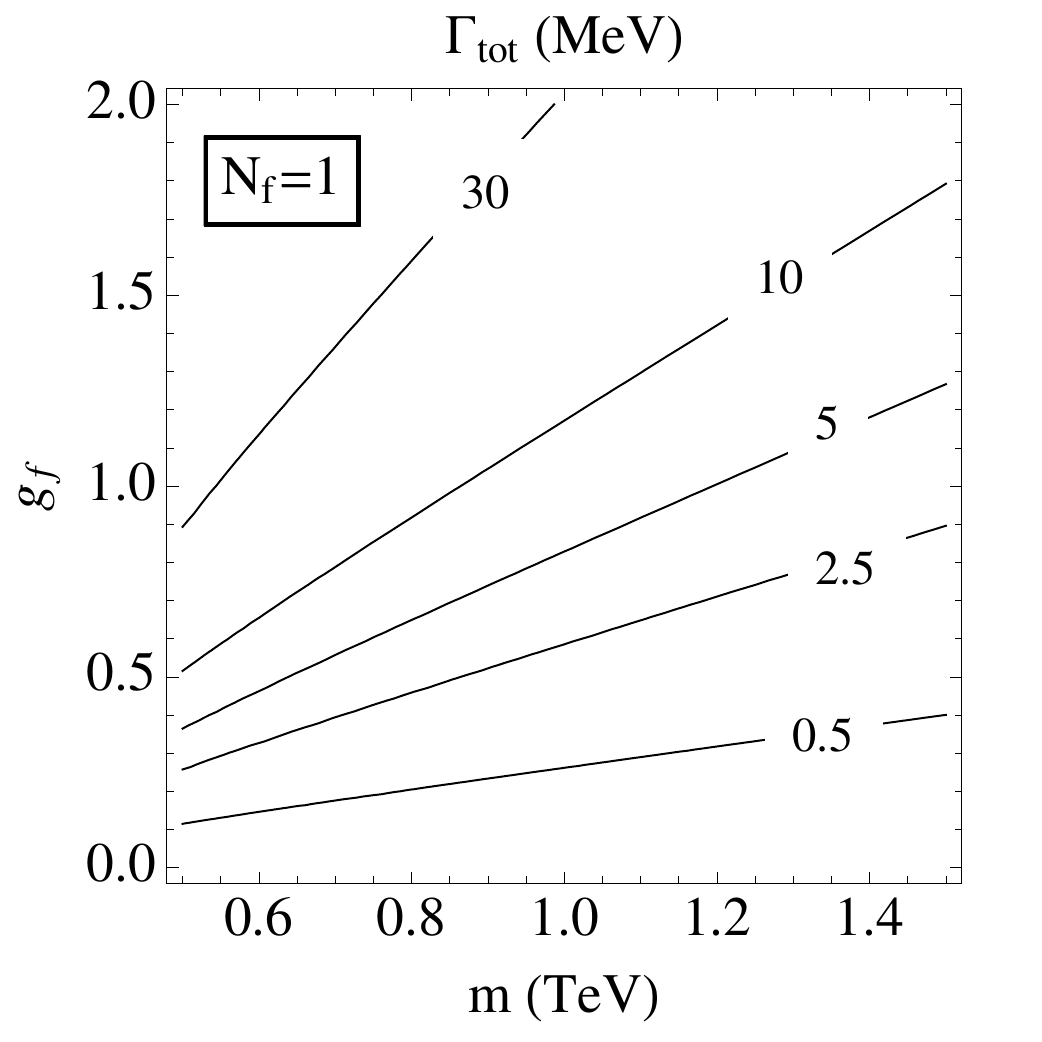}
\caption{The di-photon rate (left) and total width of $\Phi$ (right) as a function of $m$ and $g_f$ for the $F1$ model with one flavor.\label{moneyplotF1}}
\end{figure}

In Fig.~\ref{moneyplotF1} we show the di-photon rate as well as the total width of $\Phi$ as a function of $m$ and $g_{f}$ for the $F1$ model with $N_f=1$. For the remaining fermionic models we do not present plots, but instead we include $R_{\gamma\gamma}$, $\Gamma_{\gamma\gamma}$ and the total width ($\Gamma_{tot}$) for an example point in table \ref{TripletBranchingRatios}. Their values over the remainder of the $(m,g_f,N_f)$ parameter space can be easily obtained by making use of the parametric scaling of these quantities:
\begin{align}
R_{ \gamma\gamma}\,\approx\, &g_f^2\times N_f^2\times  \left(\frac{m}{1\,\text{TeV}}\right)^{-2} \times R^0_{ \gamma\gamma}
\label{eq:sec2arate}
\end{align}
where $R^0_{ \gamma\gamma}$, is the benchmark point in table \ref{TripletBranchingRatios}. $\Gamma_{\gamma\gamma}$ and $\Gamma_{tot}$ scale the same way.
For completeness, the well-known full expressions for the widths are
\begin{align}
\Gamma_{ \gamma\gamma}=&\frac{\alpha^2 N_c^2 N_f^2}{1024\pi^3}\frac{m_{\Phi}^3}{m^2}\left|2 g_f \left(\sum_i \;Q_i^2\right)\, A_{1/2}\!\left(\frac{m_\Phi^2}{4 m^2}\right)+ \frac{g_s}{m} \left(\sum_i \;Q_i^2\right)\, A_{0}\!\left(\frac{m_\Phi^2}{4 m^2}\right)\right|^2\\
\Gamma_{ gg}=&\frac{\alpha_s^2 N_f^2}{512\pi^3}\frac{m_{\Phi}^3}{m^2}\left|2 g_f \, A_{1/2}\!\left(\frac{m_\Phi^2}{4 m^2}\right)+ \frac{g_s}{m} \, A_{0}\!\left(\frac{m_\Phi^2}{4 m^2}\right)\right|^2,
\label{eq:gamgamgam}
\end{align}
and where the $A_{0,1/2}$ are the usual loop functions (see for instance \cite{Djouadi:2005gi}). For $m \gtrsim m_{\Phi}/2$ they can be approximated by $A_{0}\approx -1/3$ and $A_{1/2}\approx 4/3$.

For $F1$ we see in Table~\ref{TripletBranchingRatios} that a di-photon rate of several fb can be accommodated easily. For the models with smaller electric charges, like $F5$ and $F6$, $N_f$ must be at least 5. For $F7$ we need $N_f>10$ and a rather low $m$, which implies that this model is not very plausible. 

\begin{figure}
\includegraphics[width=0.4\textwidth]{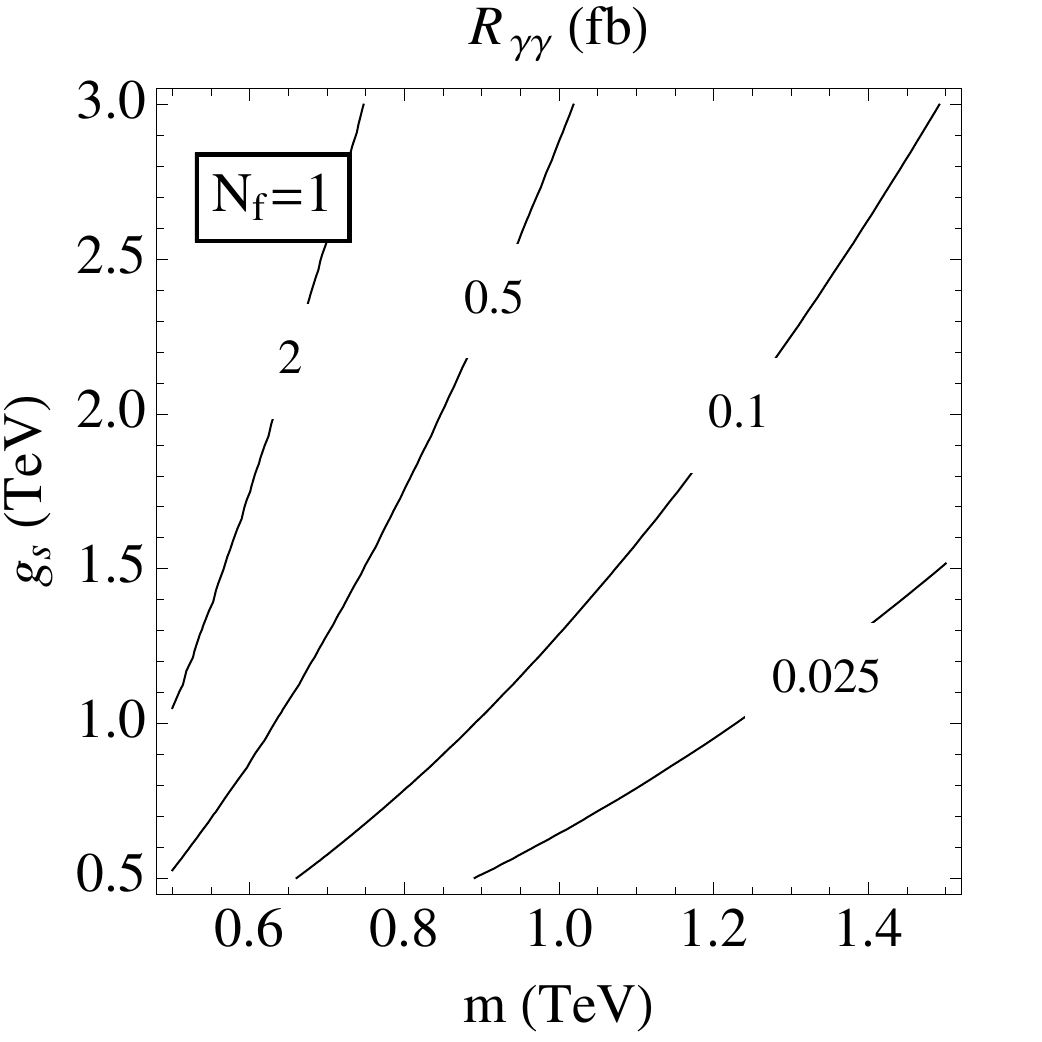}\hspace{1cm}
\includegraphics[width=0.4\textwidth]{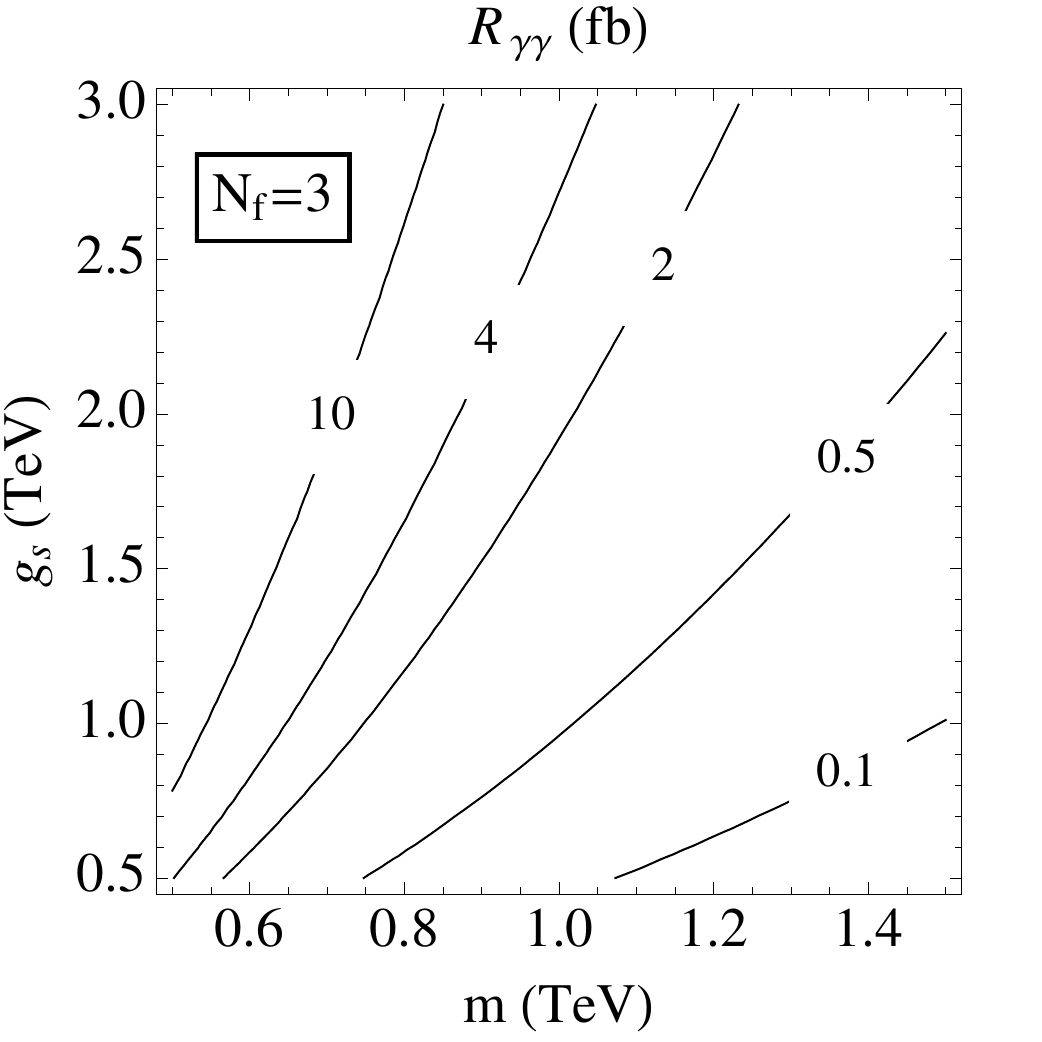}
\caption{The di-photon rate as a function of $m$ and $g_s$ for the $S3$ model.\label{fig:singletmess}}
\end{figure}

The contribution to the width from the scalar messengers is suppressed relative to that from fermionic messengers by 
$
\left(\frac{A_0}{2A_{12}}\right)^2\approx\frac{1}{64},
$
 and it is therefore difficult to obtain a large enough width in this case. $\Gamma_{\gamma\gamma}$ is largest in model $S3$, but even in this case either a very large (nearly non-perturbative) coupling or multiple flavors are needed, as shown in Fig.~\ref{fig:singletmess}.
  
Some of the models, such as $S5$,  $S6$,  $S8$,  $F4$,  $F6$ are in slight tension with the experimental upper bound on $\Gamma_{\ga Z}/\Gamma_{\ga\ga}$ ratio, and in those cases a signal in $Z\gamma$ should be observable soon.
 While our discussion here has been in terms of simplified models, more complete models are likely to contain multiple representations and in this sense a larger number of messengers may actually be very well motivated. Our results allow one to easily interpolate between models with a variety of matter content.  We will explore this in more detail in the next section.

\subsection{An Uncolored Messenger Multiplet}
\label{subsec:uncolored}

\begin{table}[t]
\begin{eqnarray*}
\begin{array}{|c|c|c|c|c|c|c|c|c|c|}
\hline
\text{Model} & \text{Representation} &  \text{$\gamma $Z/$\gamma \gamma $} &
   \text{WW/$\gamma \gamma $}&
   \text{ZZ/$\gamma \gamma $} &
   q\bar{q}\text{/$\gamma \gamma $} &R^0_{\gamma\gamma} \text{ [fb]} & \Gamma_{tot} \text{ [MeV]} & \Gamma_{\gamma\gamma} \text{ [MeV]} &
 \mbox{Decay mode} \\ \hline \hline
   \multicolumn{10}{|c|}{{\rm Scalars}} \\ \hline \hline
 \text{S11} & (1,1,-2) & 0.6 & 0. & 0.09 & 1.1\times 10^4 & 9.\times 10^{-3} & 17.9 &
   1.7\times 10^{-3} & \text{2 }e^c \\
 \text{S12} & (1,3,1) & 0.33 & 6.05 & 2.31 & 6.9\times 10^3 & 0.01 & 17.9 & 2.6\times
   10^{-3} &2\,\ell \\
 \text{S13} & \left(1,2,-\frac{1}{2}\right) & 0.82 & 9.45 & 3.32 & 1.7\times 10^5 &
   5.6\times 10^{-4} & 17.9 & 1.1\times 10^{-4} & d^c\text{ + q} \\
 \text{S14} & \left(1,2,\frac{1}{2}\right) & 0.82 & 9.45 & 3.32 & 1.7\times 10^5 &
   5.6\times 10^{-4} & 17.9 & 1.1\times 10^{-4} & u^c\text{ + q} \\
\hline\hline
   \multicolumn{10}{|c|}{{\rm Fermions}} \\ \hline \hline
 \text{F8} & (1,1,1) & 0.6 & 0. & 0.09 & 2.9\times 10^3 & 0.03 & 17.9 & 6.2\times
   10^{-3} & \ell \text{ + V/h} \\
 \text{F9} & \left(1,2,-\frac{3}{2}\right) & 0.19 & 0.38 & 0.36 & 1.2\times 10^2 & 0.81 &
   18.2 & 0.15 & e^c\text{ + V/h} \\
 \text{F10} & (1,3,1) & 0.33 & 6.05 & 2.31 & 1.2\times 10^2 & 0.79 & 19.3 & 0.15 &
   \ell \text{ + V/h}  \\
 \text{F11} & \left(1,2,-\frac{1}{2}\right) & 0.82 & 9.45 & 3.32 & 2.9\times 10^3 & 0.03
   & 18.0 & 6.2\times 10^{-3} & e^c\text{ + V/h} \\
 \text{F12} & (1,3,0) & 6.7 & 37.81 & 11.21 & 7.2\times 10^2 & 0.13 & 19.3 & 0.02 &
   \ell \text{ + V/h}  \\\hline
\end{array}
\end{eqnarray*}
\caption{Quantum numbers of the models we consider and their leading order branching fractions for various final states of the $\Phi$ decay.  The upper part of the table is for scalar loops ($\phi$), while the lower part is for fermion loops ($\Psi$). We include the di-photon rate ($R^0_{\Phi\rightarrow\gamma\gamma}$, in fb) and width to photons ($\Gamma_{\gam\gam}$, in MeV)  for a benchmark point with $m=g_s=1$ TeV, $g_f=1$, $y=0.02$ and $N_f=1$. Shown alongside the branching ratios are the decay modes for $\Psi$/$\phi$, where $V$ stands for $W$ or $Z$. }
\label{SingletBranchingRatios}
\end{table}

If $\Psi$ and $\phi$ do not carry color charge, they only contribute to the decay of $\Phi$ and not to its production via gluon fusion. Since SM gluon and vector boson fusion are challenging, as argued in the introduction, we choose to introduce an effective Yukawa coupling of $\Phi$ with the lowest generation quarks

\be
\mathcal{L}\supset y \, \bar{q} q \Phi.
\ee
Such a coupling can be UV completed in the context of a two Higgs doublet model, where $\Phi$ is a real component of the second Higgs doublet, which does not get a vev. For $y$ to be sufficiently large to be useful, this construction manifestly deviates from the minimal flavor violation ansatz, but flavor constraints can be avoided provided that $y$ is aligned with the standard model Yukawas.  This is, however, not difficult to achieve in models for dynamical flavor alignment, see for instance \cite{Knapen:2015hia}.

The rate is
\be
R_{ \gam\gam}= \frac{4}{9}\frac{\pi^2}{m_{\Phi}} \Gamma_{qq}\frac{\Gamma_{\gam\gam}}{\Gamma_{tot}}\frac{d\mathcal{L}^{q\bar{q}}}{d m_{\Phi}^2}\quad{\rightarrow}\quad \frac{4}{9}\frac{\pi^2}{m_{\Phi}} \Gamma_{\gam\gam} \frac{d\mathcal{L}^{q\bar{q}}}{d m_{\Phi}^2} \,,
\label{eq:qqprod}
\ee
where in the last equality we show how the coupling $y$ to quarks drops out in the limit where $\Gamma_{tot}\approx \Gamma_{qq}$. The natural width for $\Phi$ decaying into di-photons is now 0.1 MeV or smaller.  We write the decay width in terms of an effective coupling $c_\gamma$,
\begin{figure}
\includegraphics[width=0.28\textwidth]{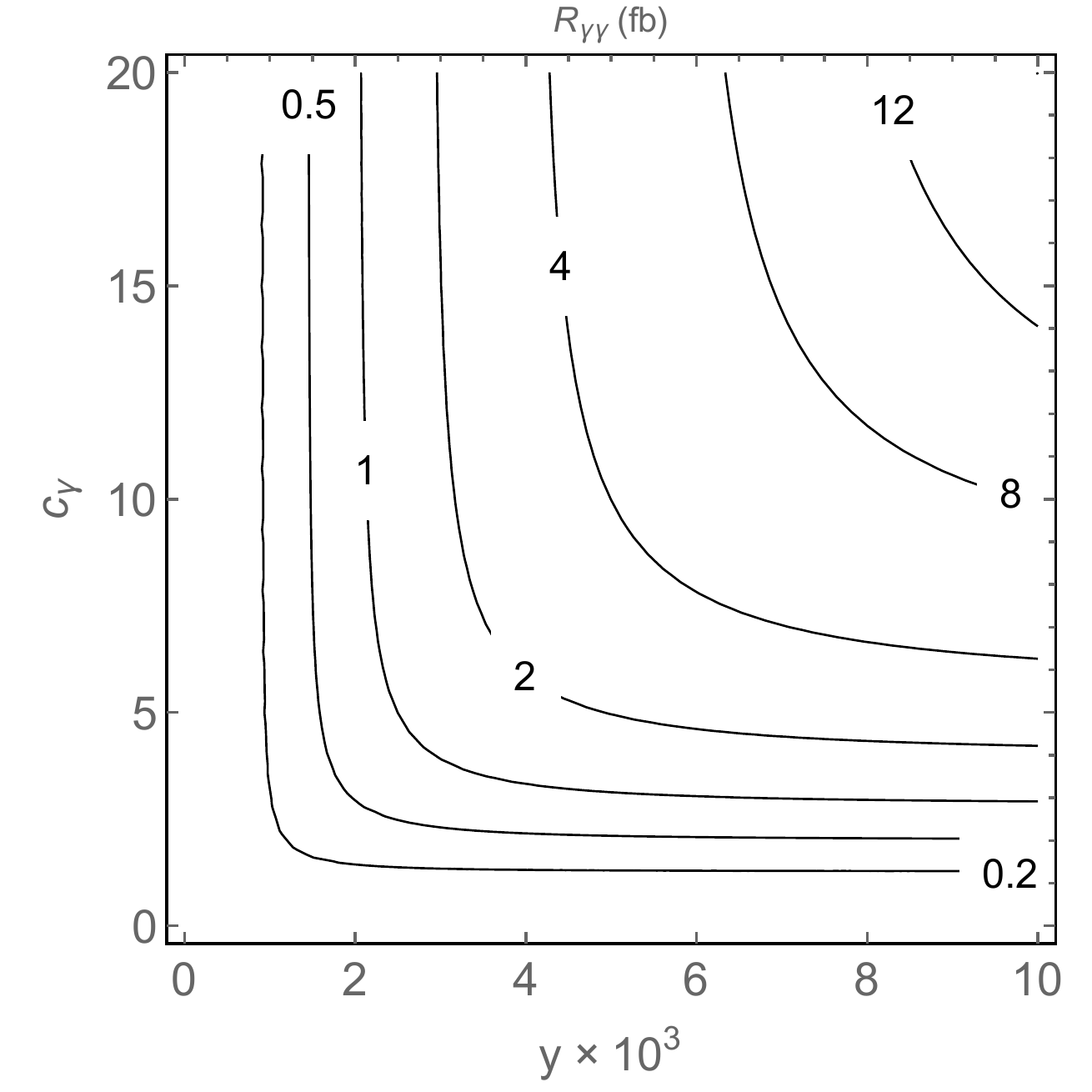}\qquad
\includegraphics[width=0.333\textwidth]{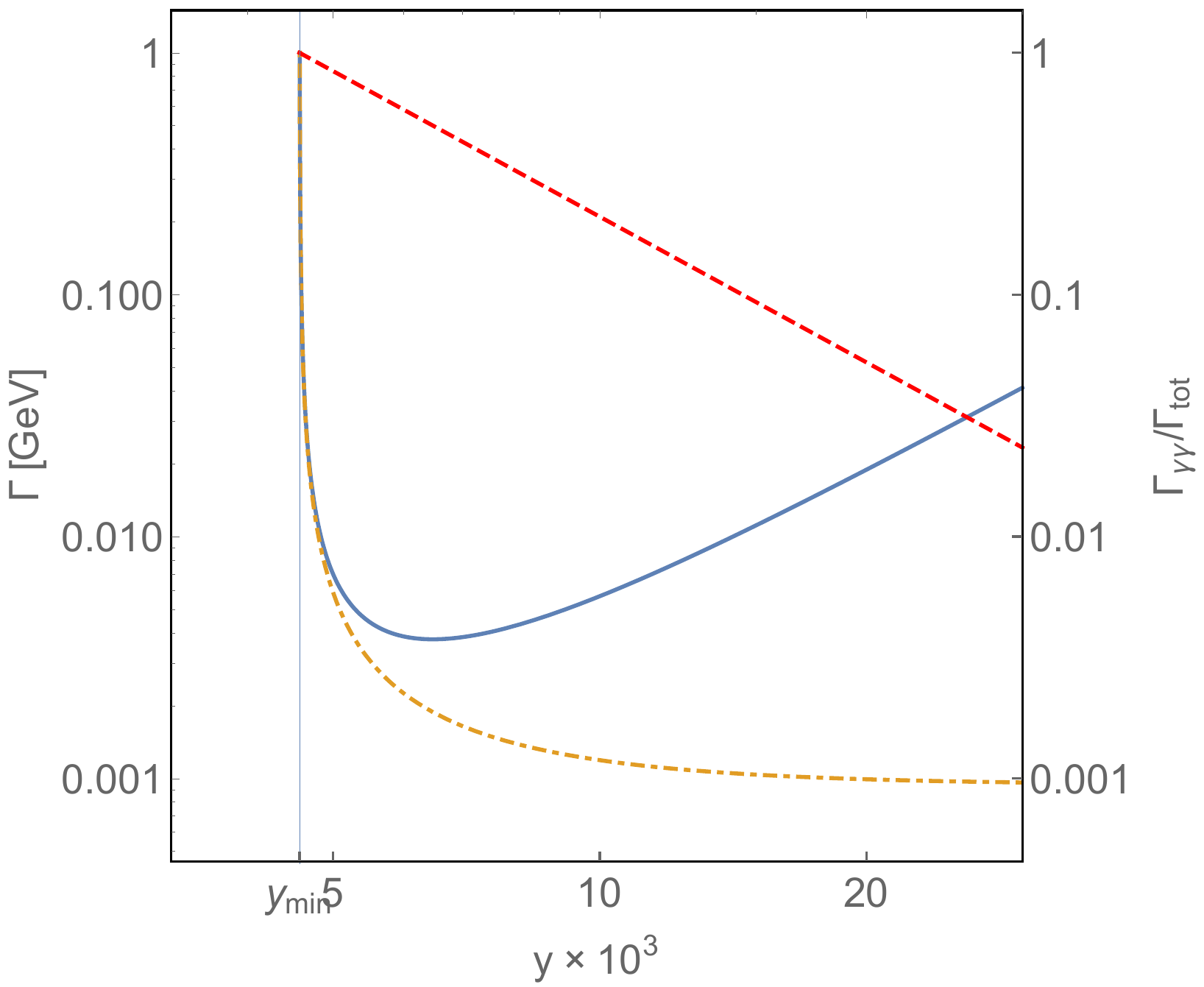}\qquad
\includegraphics[width=0.28\textwidth]{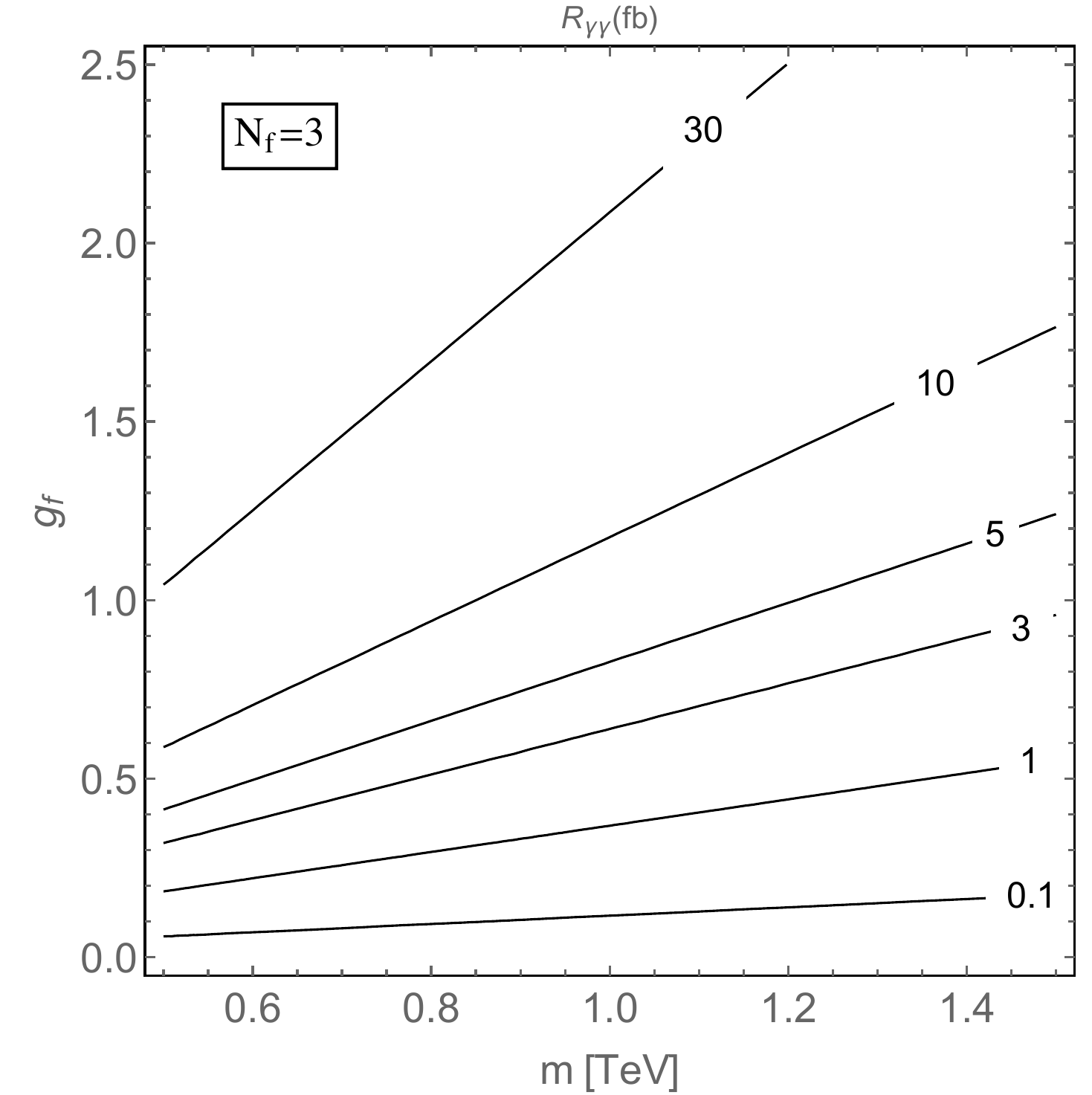}
\caption{ (a) Left: Contours of the rate to di-photons (in fb), as a function of $y$ and $c_\gam$, assuming that these partial widths dominate the total width. (b) Middle:  Fixing the rate  to the observed 5\,fb, and plotting the total decay width $\Gamma_{tot}$ (solid), the decay width to photons $\Gamma_{\gam\gam}$ (dot-dashed), both in GeV, and the ratio of these two branching fractions (dashed) as a function of $y$. (c) Right:  Contours of the rate to di-photons (in fb), for $y=0.02$ (which is large enough such that it drops out of the rate), as a function of the parameters $M$ and $g_f$ of the model $F9$ with $N_f=3$.}
\label{fig:qqplot}
\end{figure}
\be
\Gamma_{\gam\gam}=\left(\frac{\alpha}{4\pi}\right)^2 c_{\gamma}^2  \,\frac{m_{\Phi}}{4\pi}
\ee
where $c_\gamma$ can be deduced from eq.~\eqref{eq:gamgamgam}, and where
\be
\Gamma_{q\bar{q}} = N_c\frac{y^2}{16\pi}m_{\Phi}  \,.
\label{eq:2agamqq}
\ee
Repeating the exercise of the previous section, we list the possible representations for $\Phi$ that we find in Table~\ref{SingletBranchingRatios}. In Fig.~\ref{fig:qqplot}(a) we plot the rate to di-photons as a function of the Yukawa coupling and the effective coupling, $c_\gam$, for the benchmark point of the $F9$ model. As the Yukawa coupling becomes large, we see the effect of it dropping out of the rate---contours of constant $c_\gam$ become horizontal. This can also be seen in Fig.~\ref{fig:qqplot}(b) where the total width and the partial width are shown as a function of $y$, along with their ratio. As $y$ increases from its minimal value, $y_{min}$,
 the total width increases, and it forces the di-photon partial width to plateau.

This behavior is different from the models in Sec.~\ref{sec:coloredmess}, where a single messenger coupling controlled both the production and decay. There, there was no freedom to increase the width by dialing the coupling, since the rate provided an anchor. Here, on the other hand, the width can be increased by increasing $y$, and the rate to photons stays constant. Of course, this also increases the production cross section, and, as shown in the left panel of Fig.~\ref{moneyplot}, at some point di-jet constraints place an upper limit on this coupling from the decay back to light quarks; this occurs around $y\sim 0.16$.   Thus, even with tree level decays to quarks, unless there are additional (exotic) decay channels of $\Phi$, it is difficult to achieve a width much larger than 1 GeV, as shown in Fig.~\ref{moneyplot}.

The option to add multiple representations is of course also open in this case; these serve to increase the width.  In contrast with a colored messenger, we find that none for the models can match the rate without in fact doing this. Fig.~\ref{fig:qqplot}(c) shows contours of the di-photon rate in terms of the mass $m$ and coupling $g$ of the messenger for the $F9$ model; for all of these models we find that we need $N_f\sim3-10$ in order to obtain the rate. Modulo group factors, this accounts for the relative missing factor of $N_c$ compared with the rates in Table~\ref{TripletBranchingRatios}, and so is to be expected. In fact, we could consider the above model with colored particles running in the loop, where the distinction with the previous section is that now also a tree-level production mechanism is open, in addition to gluon fusion.

To reiterate and summarize, the main advantage of these uncolored (or perhaps more pertinently $qq$ coupling) models over the colored messenger model is an extra parameter that provides some freedom, {\em e.g.}, to increase the particle's width if one does not want to add extra messengers.  This is not enough, however, to make the particle 10's of GeV broad without introducing exotic decays (although the messengers may provide such decays if their mass is sufficiently light).

In the next section we discuss how to more cleanly separate production from the branching fractions of $\Phi$ via a cascade. 

\section{Di-photons from Cascades}
\label{sec:cascades}

In this section we discuss how the di-photon signal can originate from the decay of a heavier messenger, $M$. 
Such a topology implies that the event should contain extra structure, in addition to the di-photon resonance, such as extra SM resonances or energetic jets.  It also implies that $\Phi$ may be moderately boosted.  The apparent absence of significant extra activity in the di-photon events points towards lighter resonances, not far from 1~TeV. However it also provides an opportunity for discovery of additional states.
As discussed in Sec.~\ref{sec:width}, the cascade decay topology does not completely alleviate the minimum requirements on $BR(\Phi\rightarrow\ga\ga)$, as was shown in Fig.~\ref{moneyplotCascade}.  But, it does allow us to suppress the branching ratio in SM final states listed at the beginning of Sec.~\ref{sec:intro}, while maintaining a sizable production rate for a $\Phi$ with a substantial width. This is possible at the price of:
\begin{itemize}
\item Becoming sensitive to other LHC BSM searches: now what is relevant is not just the decay products of $\Phi$ alone, but the full final state of the $M$ decay. Some of the searches listed in the introduction are sufficiently inclusive and in principle can still be sensitive even in the presence of additional activity in the event from the messenger decay.  They thus still pose a constraint on the $\Phi$ branching ratios.  However, more exclusive searches may provide even more stringent bounds. For example $W^{\prime} \rightarrow W \Phi \rightarrow WWW$ may produce same sign di-leptons and jets, and $\Phi$'s $WW$ final state may be more constrained by same-sign di-lepton searches than from direct $WW$ resonance searches. Establishing whether certain decay modes are compatible with the di-photon signals in the cascade decay case requires recasting other searches, for which a direct interpretation in the models studied here is not available.  This is left for future work. Therefore satisfying the bounds quoted here is a necessary, but not sufficient, condition for viability.
\item Introducing new states below $m_{\Phi}/2$ as a way to increase $\Gamma_{\Phi}$ via tree level decays. This opens up various possibilities: $\Phi$ can still inherit SM-like branching fractions at a sub-leading level and have a sizable total width, without having to increase the partial width to di-jets. Conversely, $\Phi$ may be secluded from the SM and accessible only through the messenger decay, while acquiring a sizable width through a new hidden sector.
\end{itemize}

There are two main possibilities for $M$ that we consider in detail:
\begin{itemize}
\item a heavier bosonic resonance, $W^{\prime}$ decaying into $\Phi + W$, with $\Phi \rightarrow \ga\ga$, and
\item a fermionic resonance produced in the $s$-channel and decaying in $Q'\rightarrow\Phi+q$ (where $q$ can also be a top or a bottom quark).
\end{itemize}
Both topologies may originate in many well-motivated models, such as composite Higgs~\cite{Panico:2015jxa,Bellazzini:2014yua,Mrazek:2011iu}, Little Higgs~\cite{ArkaniHamed:2002qx,Schmaltz:2005ky}, left-right models~\cite{Senjanovic:1975rk,Mohapatra:1974gc,Mohapatra:1974hk,Pati:1974yy}, two Higgs doublet models~\cite{Branco:2011iw,Djouadi:2005gj}, and more generally in models with extra gauge and/or heavy quark partners at the TeV scale~\cite{Serra:2015xfa,Redi:2013eaa,Matsedonskyi:2015dns}. We will discuss some of the possibilities below.  Of course, it is also possible that $M$ is a scalar resonance.

Rather than to analyze  a specific example model in detail, we instead focus on collecting quantitative information about the phenomenological requirements on the decay widths of the heavier resonance $M$ and of $\Phi$ that allow for the observed $\ga\ga$ rate without violating the most relevant past LHC Run I searches. This information is central to determining how specific models can be accommodated by the observation, as well as the most promising channels in which to search.

We will express the results as a function of the mass of the heavier messenger resonance, $M_{M}$, $M=W^{\prime},Q^{\prime}$, and quote the limits on branching ratios such as $BR(W^{\prime} \rightarrow WZ)/BR(W^{\prime} \rightarrow W\Phi)BR(\Phi \rightarrow \ga\ga)$. We hereby consider a broad range of final states, as enumerated in Table~\ref{tab:ResonanceInputs}.  The results are shown in Fig.~\ref{fig:cascades}.  
For the case of a fermionic resonance, the situation is a bit more complicated by the fact that, being colored, most of the searches have been performed in the pair production topology. These searches set upper limits on branching ratios of the fermionic resonance into a variety of final states, but most of them are ineffective above 750-800~GeV. We refer the reader to the experimental results for those that are able to set a limit between 800 and 950~GeV~\cite{Khachatryan:2015oba,Chatrchyan:2013oba,CMS:2014dka,CMS:2014bfa,Aad:2015kqa,Aad:2015mba}.
The only single production searches performed at Run I are for a $B^{\prime}$ resonance, produced from a $bg$ initial state and decaying to either $bg$ or $tW$ (and more generally $jj$), and we will therefore quote the results for those decay modes. 

\begin{figure}
\includegraphics[width=0.42\textwidth]{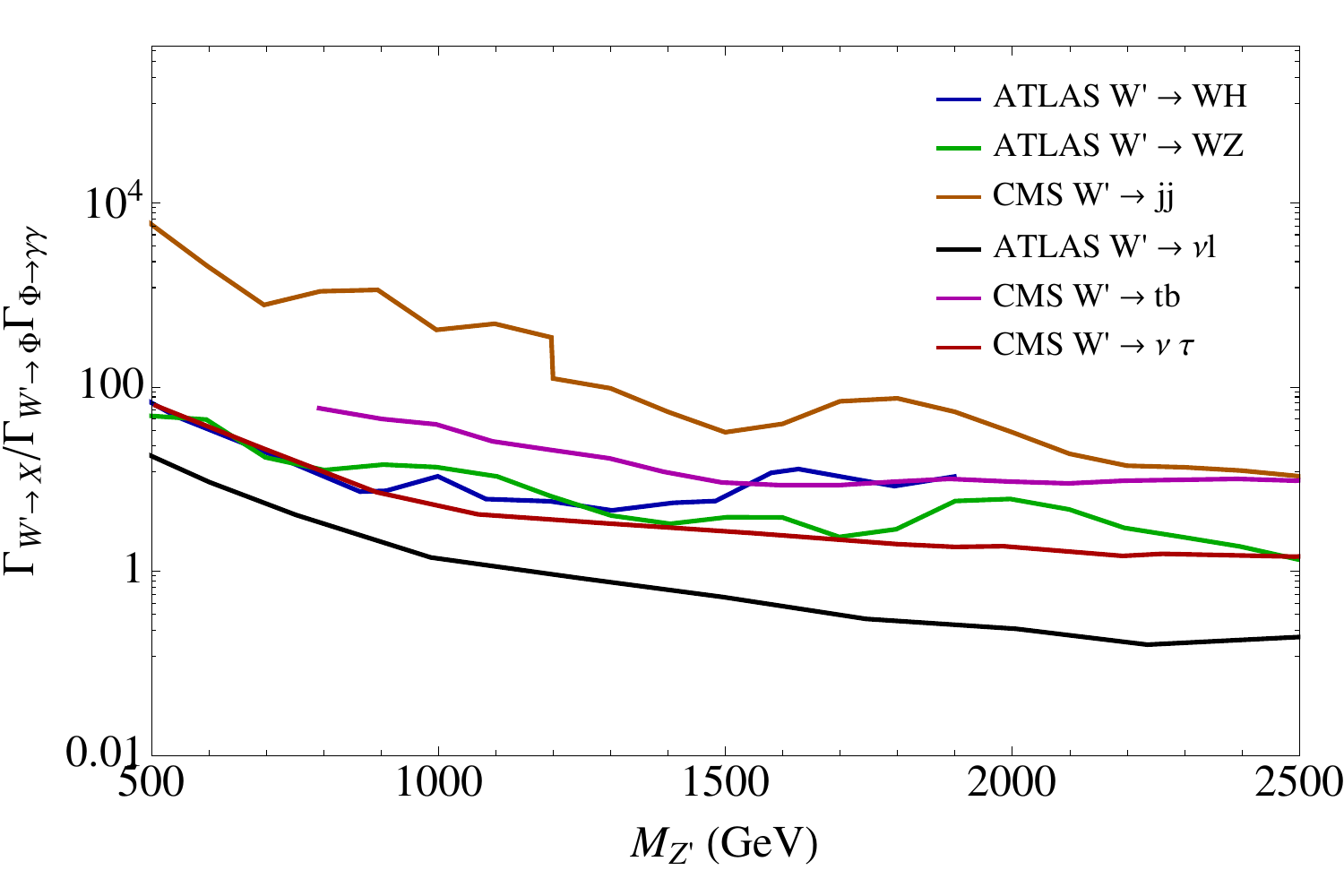}\qquad
\includegraphics[width=0.42\textwidth]{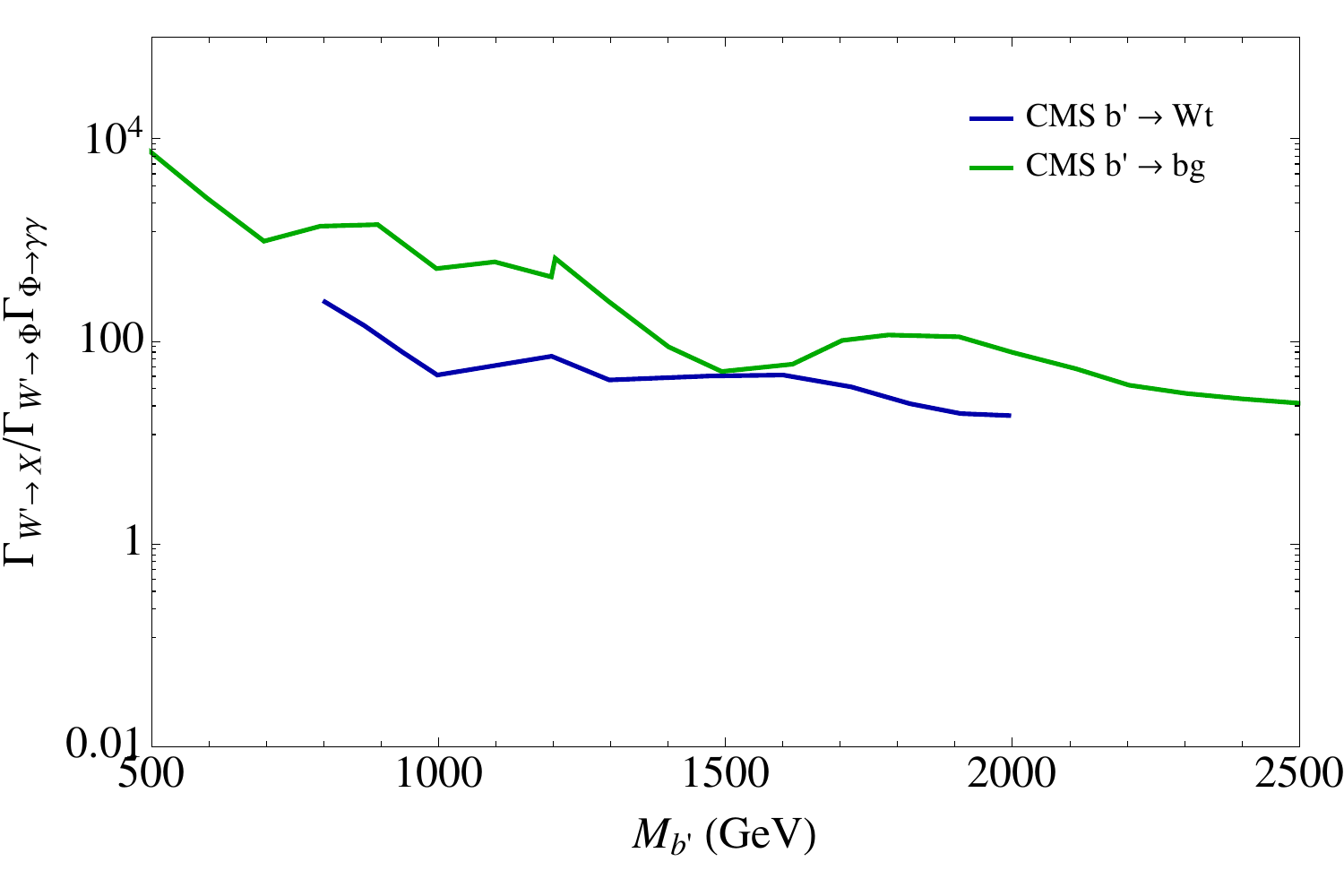}\qquad
\caption{Upper limits on branching ratios for messenger decays.\label{fig:cascades}}
\label{fig:cascades}
\end{figure}

The constraint on the product of branching ratios $BR(M\rightarrow \Phi X)BR(M\rightarrow jj)BR(\Phi\rightarrow \ga\ga)$ can be extracted from Fig.~\ref{moneyplotCascade} by rescaling the curves by $BR(M\rightarrow \Phi X)BR(M\rightarrow jj)\times4$.
While the constraints from other channels must be taken into account, one can see from Fig.~\ref{fig:cascades} how they may be satisfied. Note from these figures that for heavier resonances both partial widths $\Gamma(M\rightarrow \Phi X)$ and $\Gamma(\Phi\rightarrow \ga\ga)$ should increase to maintain a large enough rate, and it becomes increasingly difficult to satisfy the $\Phi\rightarrow\ga\ga$ requirements.

\begin{table}
\begin{center}
\begin{tabular}{c||c}
\hline
Search & Process \\
\hline
\hline
1503.08089~\cite{Aad:2015yza} & $W^{\prime}\rightarrow WH$ \\
\hline
ATLAS-CONF-2015-045~\cite{ATLAS-CONF-2015-045} & $W^{\prime}\rightarrow WZ$ \\
\hline
1407.7494~\cite{ATLAS:2014wra} & $W^{\prime}\rightarrow \ell \nu$ \\
\hline
1402.2176~\cite{Chatrchyan:2014koa} & $W^{\prime}\rightarrow tb$ \\
\hline
1508.04308~\cite{Khachatryan:2015pua} & $W^{\prime}\rightarrow \tau\nu$ \\
\hline
1501.04198~\cite{Khachatryan:2015sja} & $W^{\prime}\rightarrow jj$ \\
\hline
CMS-PAS-EXO-14-005~\cite{CMS:2015neg} & $W^{\prime}\rightarrow jj$ \\
\hline
\end{tabular}
\end{center}
\caption{Searches utilized in Fig.~\ref{fig:cascades} to constrain production of $\Phi$ via cascade decays. \label{tab:ResonanceInputs}}
\end{table}

We are now at the stage where we can speculate about the nature of the $\Phi$ and $M$ resonances. Given the early stage, here we provide only some broad options leaving detailed quantitative studies for the future. $\Phi$ could be a Higgs boson. For example in a left-right model it may be the neutral component of the $SU(2)_{R}$ triplet $\Delta$ responsible for breaking $SU(2)_{R}$. In this model there are naturally additional charged states that can boost the $\ga\ga$ rate: two more scalars of charge 1, a scalar of charge 2 and a charge-1 vector, the $W^{\prime}$ that can also act as the messenger $M$. If we assume ${\cal O}(1)$ couplings of these states to $\Phi$, and no large scale separation in their masses, their presence can account to ${\cal O}(20)$ effective flavors of charge-1 scalars. This is sufficient to drive the di-photon partial width in the ${\cal O}(10)$~MeV range. Furthermore $\Phi$ can also mix with the SM Higgs at ${\cal O}(v/v_{R})\simeq (g_{R}/g_{L}) (m_{W}/m_{W^{\prime}})$, inducing decays to SM particles at a safe level. The $W^{\prime}\rightarrow \Phi W$ branching ratio is impacted by mixing angles naturally of a similar order as the one controlling that to di-bosons. Decays into $tb$ and di-jets tend to dominate over decays to light di-bosons, but a $BR(W^{\prime}\rightarrow\Phi W)$  at ${\cal O}({\rm few}\,\%)$ can still be naturally achieved.  For example $BR(W^{\prime}\rightarrow \Phi W)\sim 0.05$, $BR(\Phi \rightarrow \ga\ga)\sim 30\,{\rm MeV}/30\,{\rm GeV} = 10^{-3}$ would work for $M_{W^{\prime}} \sim 1.5\,{\rm TeV}$. $\Phi$-Higgs mixings would be at the 5\% level, consistent with Higgs properties and bounds on $\Phi$ direct searches. Of course, this mixing is not sufficient to generate a $\sim 45$  GeV width, so extra decay channels are necessary (which, in SUSY left-right models, may come from tree-level decays into the electroweak-ino sector).  More detailed studies are needed to render this estimate on firmer grounds.

In composite Higgs models, depending on the global symmetry group of the strongly coupled sector, one may find extra scalar states as pseudo-Nambu-Goldstone-Bosons (pNGBs) that can fill the role of $\Phi$. Furthermore, the presence of heavier resonances, both bosonic and fermionic (in the form of top partners), may provide natural candidates for the messenger $M$.  At the same time, if charged, they contribute to $\Phi$'s di-photon width. Extra pNGB singlets may be somewhat heavier that the weak scale~\cite{Gripaios:2009pe,Sanz:2015sua} but still lower that the composite scale, and thus relatively narrow. Depending on the properties of the UV completion, the di-photon rate may also be enhanced by Wess-Zumino-Witten terms. Unfortunately, for the most economical composite Higgs model containing an extra singlet $SO(6)/SO(5)$, this is not the case~\cite{Arbey:2015exa}.  And, while a WZW term is present, it vanishes exactly for the photons, rendering the di-photon width of this extra singlet negligible. Nevertheless, other constructions may allow this possibility. 

Other possibilities for $\Phi$ and $M$ may be found in extended Higgs sectors where $\Phi,\,M$ may be extra Higgs bosons with reduced contribution to the EWSB (alignment-limit) and with significant couplings to new non-SM states (such as extra fermions, possibly Dark Matter).

We conclude this section with the discussion of a different cascade topology that can give rise to a di-photon signal, albeit without the presence of a resonance decaying to $\ga\ga$. The di-photon signal can be a kinematic edge in the cascade decay of 
\be
A\rightarrow \ga (B\rightarrow\ga C)
\ee
which is currently being misinterpreted as a peak due to the low statistics. In that case, as is well known, we have a relation between the A, B and C masses given by
\be
m_{\ga\ga,max}^{2} = \frac{(m_{A}^{2}-m_{B}^{2})(m_{B}^{2}-m_{C}^{2})}{m_{B}^{2}},\label{eq:edge}
\ee
and, for a given final state C,  there are resonant peaks in the $\ga C$ and $\ga\ga C$ invariant mass distributions, which at the moment may not be visible yet due to the low statistics and/or combinatoric backgrounds. $A$ and $B$ will necessarily be new, non-SM particles, while $C$ can be a SM state. Two interesting cases for $C$ are a vector boson, either a W or a Z boson, or a jet\footnote{The case of a top or bottom quark or a Higgs boson would imply the presence of at least one b-jet in most of the events and should be fairly easy to investigate.}.

Let us now assume that $A$ is singly produced in proton-proton collisions, which is motivated by the little activity in the rest of the signal events. Similar to Eq.~(\ref{BrBrBr}), fixing the di-photon rate in terms of $A$ and $B$ branching ratios and masses, we then have
\be
\frac{\Gamma_{A}}{M_{A}}\left(\frac{d\mathcal{L}}{d M_{A}^2}  c \right) BR(A\rightarrow jj)BR(A\rightarrow B\ga)BR(B\rightarrow \ga C) = 10\,{\rm fb}.\label{eq:edge-rate}
\ee
The only information that can be inferred from this equation is a lower bound on $BR(A\rightarrow jj)BR(A\rightarrow B\ga)BR(B\rightarrow \ga C)$ as a function of $M_{A}$, which trivially states that if these branching ratios are too small, one cannot accommodate the observed di-photon rate.
One cannot make further progress without knowing more information on the nature of $A$ and/or $B$. If $B$ cannot be directly produced in proton proton collisions,  one presently can easily explain the bump as long as Eqs.~(\ref{eq:edge},~\ref{eq:edge-rate}) are satisfied. Future observations of the $\ga C$ and $\ga\ga C$ peaks or the $\ga\ga$ line-shape are the only handles to disprove this possibility.

On the other hand, in models where $B$ can also be produced directly in proton collisions, further constraints apply: ATLAS and CMS have performed searches for $B\rightarrow \ga X$ for $X=W,Z,j$, besides di-jet searches. We can use these searches to set upper limits on $BR(B\rightarrow \ga C)\times BR(B\rightarrow j j)$ and on $BR(B\rightarrow j j )$, with the same techniques employed above. 
At the same time, by using Eqs.~(\ref{eq:edge},~\ref{eq:edge-rate}), and using the fact that $BR(A\rightarrow jj)BR(A\rightarrow B\ga)<1/4$ we can extract a lower limit on $BR(B\rightarrow \ga C)$.
The results are shown in Fig.~\ref{fig:Edge}, where the upper limits from direct searches are expressed as solid lines and the lower limit from the requirement to have enough rate to fit the excess is expressed as a dashed line. Satisfying both constraints is equivalent to imposing an upper bound on $BR(B\rightarrow jj)$ which is non-trivial only for low enough $M_{B} \lesssim 600$~GeV in the case $B\rightarrow Z\ga$.
\begin{figure}
\includegraphics[width=0.4\textwidth]{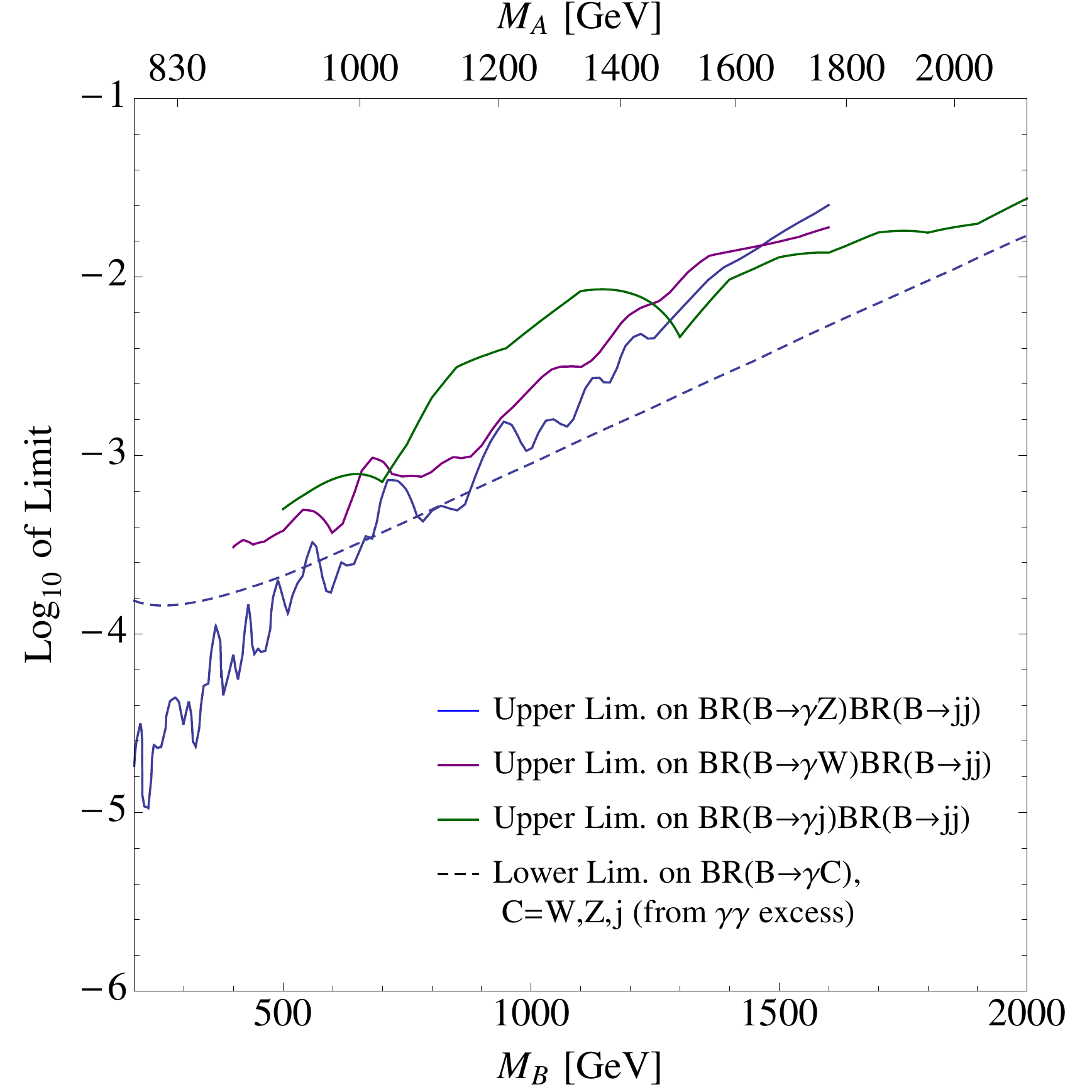}
\caption{Summary of the upper limits on the branching ratios of the intermediate particle $B$ in the edge topology (last case of Table~\ref{tab:topologies}) into $W\ga$, $Z\ga$, $j\ga$ times the branching ratio into di-jets from direct searches at Run I. The lower limit on the branching ratios set by the required di-photon rate is shown as a dashed line. The combination of these curves sets an upper limit on $B\rightarrow jj$. For each mass $m_{B}$, the corresponding mass $m_{A}$ of the parent particle required to produce an edge at 750~GeV is shown on the top edge of the frame.\label{fig:Edge}}
\label{fig:Edge}
\end{figure}
We conclude that there are no obstructions from Run I searches for explaining the di-photon rate with a kinematic edge. However the absence of significant extra activity in the events points towards the presence of the intermediate state $B$ not too far from 500~GeV.

\section{Collimated pairs of photons? Hidden valley models}
\label{sec:HV}

In the previous section we discussed models where $\Phi$, being neutral, couples to two photons via loops of charged particles. The natural size for this partial width is $ \alpha_{em}^{2}/256\pi^{3}m_{\Phi}$ times the appropriate Casimirs and powers of electric charges, which puts us in the 10 keV - 1 MeV range. One can increase it by increasing the multiplicity and the charge of the particles running in the loop, as well as the strength of their coupling with $\Phi$. However it is impractical to render this partial width of the order of a tree level decay, ${\cal O}(5\,{\rm GeV})$, without going into the strong coupling regime  and/or having a very large number of flavors. Nevertheless, from the discussion in Sec.~\ref{sec:intro}, there are currently no obstructions from the LHC data for $\Phi$ to have a tree-level size partial width into what looks like a pair of photons. In this Section we investigate whether this can be achieved if the two photons recorded in the experiments are actually highly collimated pairs of photons. The possibility for such photon-jets has previously been considered in detail, and in particular as an exotic decay mode of the standard model Higgs \cite{Draper:2012xt,Ellis:2012zp,Ellis:2012sd,Dobrescu:2000jt,Toro:2012sv,Chang:2006bw,Curtin:2013fra}. 

The obvious class of models producing this topology are Hidden Valleys~\cite{Strassler:2006im,Strassler:2006ri}, in which $\Phi$ decays at tree level into a pair of light scalars, $\phi$, which in turn decays to a pair of photons. The collimation is achieved by requiring that $m_{\phi}$ is sufficiently low, below a GeV. The coupling of $\phi$ to photons can be generated by loops of massive charged fermions (denoted below by $\psi$). In this way we have decoupled the size of the $\Phi$ width to the necessary requirements of loop-mediated $\ga\ga$ dictated by gauge invariance. In this class of models $\Phi$ can, in principle, be a scalar or a vector (or a tensor), but we will focus on the scalar hypothesis below, as it is simplest to embed in a model. 
We consider the following simplified model:
\be
{\cal L} \supset  |\partial_{\mu}\Phi|^{2} + \frac{1}{2} m_{\Phi}^{2}  \Phi^{2} +\frac{1}{2} \left(\partial_{\mu}\phi\right)^{2} +\frac{1}{2}(m_{\phi}^{2} + g m_{\Phi} \Phi)\phi^{2} + i\bar{\psi}\slashed{D}\psi -m_{\psi}\bar{\psi}\psi +y \phi \bar{\psi}\psi 
\ee
where we have normalized the trilinear scalar interaction to the $\Phi$ mass. We note that, since the decay now proceeds at tree level, in principle $\Phi$ can be identified with a heavy Higgs  in an extended Higgs sector such as  the 2HDM as long as $BR(\Phi\rightarrow \phi\phi) \gtrsim {\cal O}(1\%)$. 
The next constraint is on the $\phi$ mass, due to the photon pair collimation. We know from Higgs measurements that the LHC experiments are able to distinguish photons of $m_h/2 \sim 65$~GeV from $\pi^{0}$'s of the same energy. This sets an upper limit on the $\phi$ mass $m_{\phi} < m_{\pi^{0}} m_{\Phi} / m_{h} \simeq 800\,{\rm MeV}$. Furthermore the requirement $2m_{\psi}\gg m_{\phi}$, to ensure a sizable decay of $\phi$ to two photons via a loop of charged fermions, will generally render the decay displaced
\be
\tau_{\phi} \simeq 3 \mbox{ mm} \left(\frac{600 \mbox{ MeV}}{m_{\phi}}\right) \left(\frac{m_\psi}{100 \mbox{ GeV}}\right)^2 y^{-2}, 
\label{philife}
\ee
further affecting the differences between the electromagnetic shower shapes between a $\phi$ decay and a prompt photon of the same energy. 
In order to asses the differences we estimated the discrepancies in the shower shapes between a photon and a displaced $\phi$ decay of 375~GeV of energy using the procedure described in Appendix~\ref{app:shower}.  

Since a prompt decay prefers a heavier mass of $\phi$, opening a decay via $G_{\mu\nu}G^{\mu\nu}$ to pions,  a requirement of a significant branching to $\ga\ga$ (in order not to suppress the total rate) is now doubly important for $\phi$, forcing us to discard the option to couple $\phi$ to the SM fermions proportionally to their masses. Therefore we are left with two options: either to couple $\phi$ to the $\tau$ lepton but not to muons or electrons, or  to assume that $\psi$ is a new charged vector-like fermion at the weak scale. The former case originates naturally in models such as those described in~\cite{Knapen:2015hia}.

\begin{figure}
\includegraphics[width=0.4\textwidth]{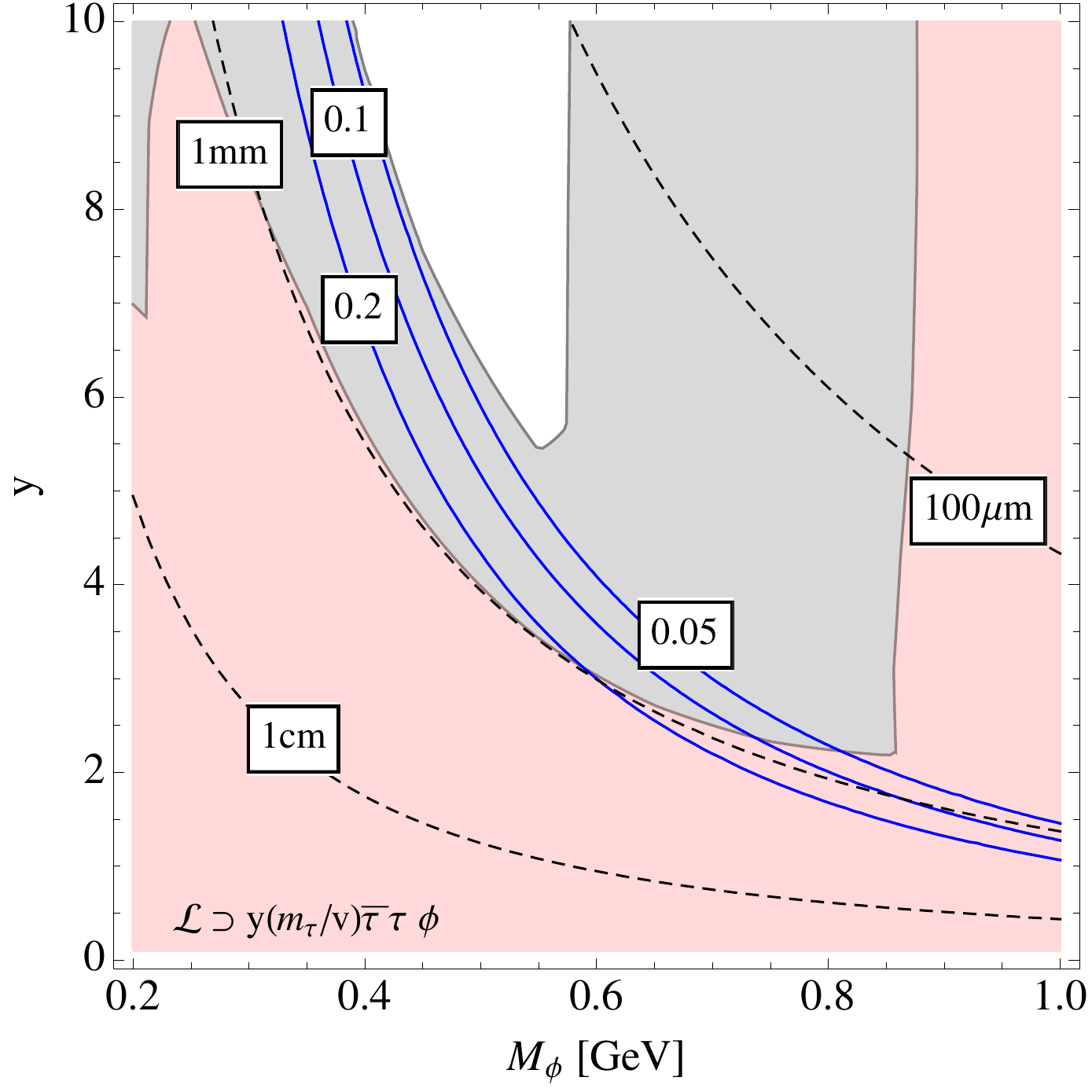}\hspace{1cm}
\includegraphics[width=0.4\textwidth]{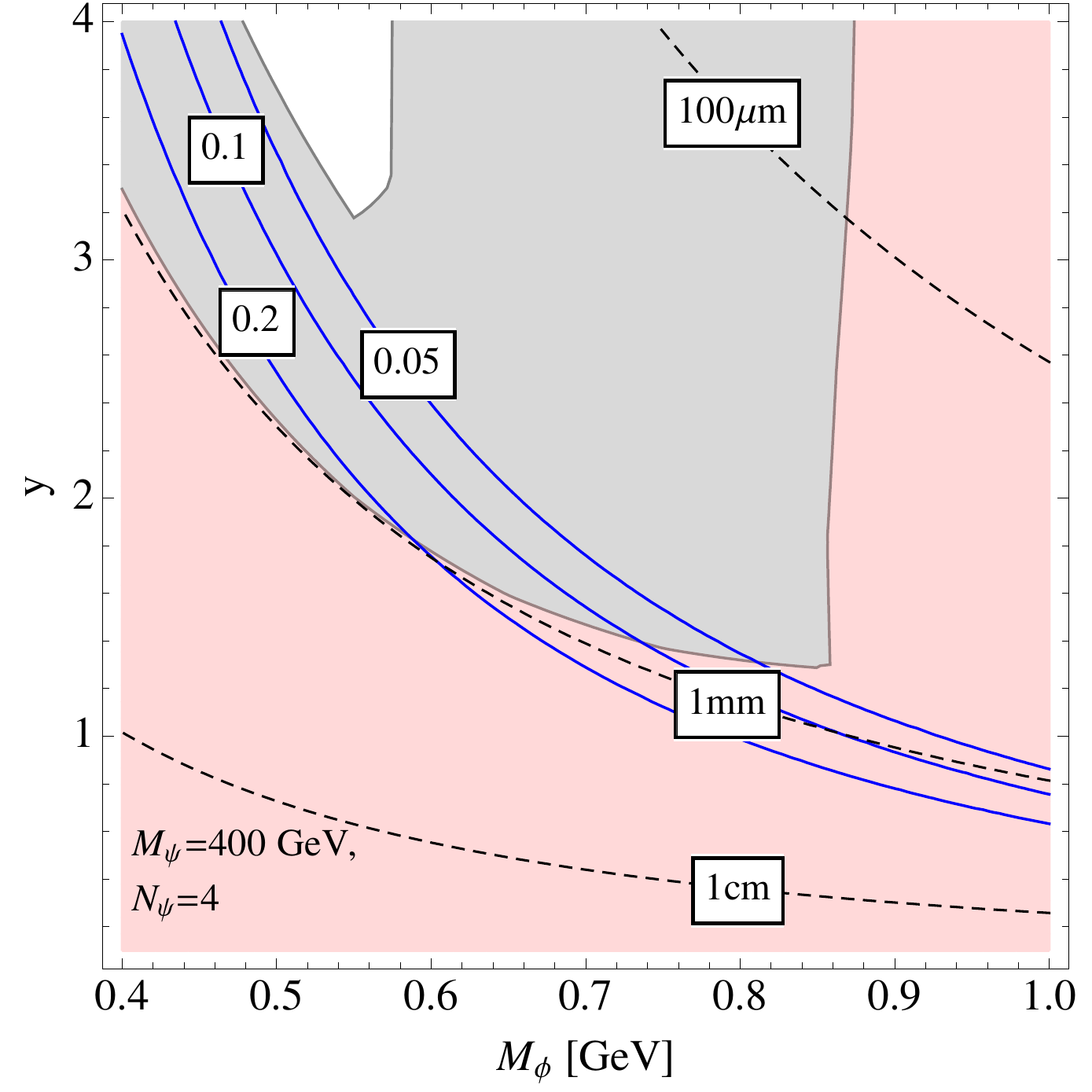}
\caption{Allowed regions for the case of $\Phi$ decaying into an Hidden Valley as a function of the mass of the light particle $\phi$ and its coupling $y$ to a charged particle generating the $\ga\ga$ partial width. $\phi$ proper decay lengths are shown as dashed lines. Blue lines describe the fraction of $\phi$ particles decaying inside the calorimeter (5,10 and 20\% respectively), while the white, red and gray areas correspond, respectively, to the degree of collimation the two photons from $\phi$ decay: indistinguishable from a single photon (according to the way of estimating the shower shape discrepancies described in Appendix~\ref{app:shower}), distinguishable (hence excluded) or the intermediate regime where further study may be needed. The left plot corresponds to $\phi$ coupling to the SM $\tau$ lepton only, with coupling $y*y_{b}$, while the right plot corresponds to the case where $\phi$ couples to 4 copies of vector-like uncolored fermions carrying unit charge with 400~GeV of mass.\label{fig:hv}}
\end{figure}

Our results are summarized in Fig.~\ref{fig:hv}. The red shaded region is excluded, since here $\phi$ generates EM-shower shapes differing from a prompt photon, by more than the difference between a 65~GeV $\pi^{0}$ and a 65~GeV $\ga$. The white region corresponds to similar differences as a would-be $\pi^{0}$ of 65~GeV of energy but mass below 50~MeV, and is therefore likely to be allowed, while the gray area correspond to the intermediate range where a more proper analysis may be needed. The blue curves denote the fraction of times $\phi$ decays inside the calorimeter. The region below the blue lines is therefore excluded by displaced searches. Finally we indicate the proper decay lengths with dashed lines.

\section{Discussion}

The possibility of a new resonance at 750 GeV is exhilarating, though, depending on the nature of these events, modeling the excess with a theory requires a bit of non-trivial structure in the new sector that should make the next run of the LHC an exciting adventure.  We summarize the possibilities and conclusions from this short paper.  
\begin{itemize}
\item As emphasized in the introduction, simply extending the SM by a single new resonance is not viable -- because the mediator is more massive than all the SM particles, tree level decays back to SM particles mediating the coupling to $\ga\ga$ gives rise to conflict with constraints, such as $t \bar t$, $W^+ W^-$ or di-jet bounds.  
\end{itemize}
The next simplest possibility is to extend the SM by two particles -- the $\gamma \gamma$ resonance as well as messenger particle(s).  We explored three possibilities along these lines.
\begin{itemize}
\item Loops of new messenger multiplets mediate the decay of the resonance.  We considered two possibilities for the production: tree-level production of the resonance by coupling to initial state quarks, or production via loops of (colored) messengers.  We explored both possibilities, and both are viable.  If the resonance is to be broad, in both cases, a relatively large 't Hooft coupling of the resonance to the messengers is needed; this feature is fully general for $2 \rightarrow 2$ processes.
\item A messenger resonance decays to the $\ga\ga$ resonance plus some other SM particle.  For example, in a $W'/Z'$ model, the decay channel is the messenger $W'/Z' \rightarrow \Phi + W/Z$.  Another possibility is the production of a messenger fermion $\Psi$ which decays, for example, to the resonance plus jet $\Psi \rightarrow \Phi + j$.  Of course, in this case, one reconstructs not only the $\ga \ga$ resonance, but also the messenger resonance.  It remains to be seen whether the data will support a second resonance.  
\item We also considered an edge topology where $A \rightarrow B \gamma \rightarrow C \ga\ga$, and we found it is consistent with constraints.
\end{itemize}
Another scenario is that the two observed photons are merged from four photons.  This leads to the last model possibility we considered:
\begin{itemize}
\item HV models, in which a heavy messenger resonance decays to a pair of very light (sub-GeV) neutral states.  As long as each of these neutral states decays to a pair of photons before reaching the calorimeter, the two photons from each decay merge into one.  The most challenging aspect of this scenario is constructing a model where the very light states decay quickly enough to $\ga\ga$.
\end{itemize}

Since the observed decay is to $\ga \ga$, one should obviously look for resonances in the $Z \ga$ and $ZZ$ channels.  Since production generically happens from a $qq,~gg$ or $qg$ initial state, di-jets is a common signature when the resonance or messenger decays back to the initial state, rather than to $\ga\ga$.  Each of these models would lead to additional new signatures:
\begin{itemize}
\item Messenger fermions and scalars could be pair produced at the LHC and decay to the states shown in Table~\ref{TripletBranchingRatios} and Table~\ref{SingletBranchingRatios}.  Many of these signatures also bear resemblance to searches for top partners in composite models.
\item A vector messenger resonance decaying to the $\Phi \rightarrow \ga\ga$ resonance plus $X$ gives rise to $WW,~WZ,~ZZ,~hW,~hZ$ possible states.  They also give rise to other signatures when the decay of the vector messenger to $\Phi Z,~\Phi W,~\Phi h$ is followed by a decay of $\Phi$ into other final states, such as $jj$.
\item A singly produced fermion messenger will decay not only to $\Phi + q$, but also $q q$, where the quarks could be any flavor, such as $t$.  This motivates, for example, $t j$ searches.
\end{itemize}

Thus we can see that any new physics associated with this di-photon resonance will likely give rise to a patchwork of signatures, which can be searched for in a wide variety of modes.  This must be understood first before satisfactory theories can be constructed.  But, ultimately, we hope that these particles will give new clues to shed light on deep outstanding questions about our Universe. 

\acknowledgements
We thank Zoltan Ligeti, Yasunori Nomura, Surjeet Rajendran and Dean Robinson for discussions. The work of SK and MP is supported in part by the LDRD program of LBNL under under DoE contract DE-AC02-05CH11231. TM and KZ are supported by U.S. DOE grant DE-AC02-05CH11231.  TM also acknowledges computational resources provided through ERC grant number 291377: ``LHCtheory''. 

\appendix 
\section{Other constraints}
\label{app:limits}
Here we summarize the relevant numerical inputs we have used in the numerical analyses performed in this paper.  The 8~TeV upper limit on various decay modes of a scalar resonance at 750~GeV are summarized in the following table, where we have used the 8~TeV values for the production cross sections of a 750~GeV SM Higgs quoted below to convert limits on $\sigma/\sigma_{SM}$ to cross sections, whenever necessary. CMS searches in the $WW$ final state use the $\ell\nu 2j$ channel, while ATLAS uses a combination of $2\ell2\nu$ and $\ell\nu2j$, for which we quote the limits for a narrow resonance. For the case of $Z\ga$ we assume an efficiency times acceptance $\epsilon\cdot A\simeq0.5$.

{\small
\begin{tabular}{| c || c | c || c | c |}
\hline
Proc. & $\sig \times BR$/pb  CMS  & ref & $\sig \times BR$/pb  ATLAS  & ref \\
\hline
$\gam\gam$ (13~TeV) & $7\times 10^{-3}$ & EXO-15-004~\cite{CMS-PAS-EXO-15-004,talk} & $5\div10 \times 10^{-3}$ & CONF-2015-081~\cite{ATLAS-CONF-2015-081}  \\
$\gam\gam$ & $1.37-2.41\times 10^{-3}$ & 1506.02301~\cite{Khachatryan:2015qba} & $2.42 \times10^{-3}$ &  1504.05511~\cite{Aad:2015mna} \\
\hline
$WW$, ggF & 0.294  & HIGG-14-008~\cite{CMS:2015lda} &  $3.7 \times 10^{-2}$ &  1509.00389~\cite{Aad:2015agg} \\
$WW$, VBF & 0.482  & HIGG-14-008~\cite{CMS:2015lda} & $2.8 \times 10^{-2}$ &  1509.00389~\cite{Aad:2015agg} \\
$ZZ$, $gg$, $\ell\ell\nu\nu$  & shape(cut) $7.74(10.2)\times 10^{-2}$ & HIGG-13-014~\cite{CMS:2013ada} & $1.9\times10^{-2}$   & 1507.05930~\cite{Aad:2015kna} \\
$ZZ$, VBF, $\ell\ell\nu\nu$ & shape(cut) $6.39(13.5)\times 10^{-2}$ & HIGG-13-014~\cite{CMS:2013ada} &  $1.6\times10^{-2}$ &  1507.05930~\cite{Aad:2015kna} \\
$ZZ$, $2\ell 2j$ & $6.29\times 10^{-2}$ & HIGG-14-007~\cite{CMS:2015mda} &  $3.7 \times 10^{-2}$ & 1409.6190~\cite{Aad:2014xka} \\
$\tau\tau$, ggF & $1.88(2.31)\times 10^{-2}$ & HIGG-14-029(-13-021)~\cite{CMS:2015mca,CMS:2013hja} & $1.2 \times 10^{-2} $  & 1409.6064~\cite{Aad:2014vgg} \\
$\tau\tau$, $bb\phi$ & $1.28(2.31)\times 10^{-2}$ & HIGG-14-029(-13-021)~\cite{CMS:2015mca,CMS:2013hja} &  $1.15 \times 10^{-2} $ & 1409.6064~\cite{Aad:2014vgg}   \\
$\tau\tau$  & 0.15 & EXO-12-046~\cite{CMS:2015ufa} & $1.02 \times 10^{-2}$  & 1502.07177~\cite{Aad:2015osa}  \\
$bb$ (ggF) & $2.2$ & EXO-14-005~\cite{CMS:2015neg} &   &  \\
$qq$ (ggF) & $2.2$ & EXO-14-005~\cite{CMS:2015neg} & 15.  &  1407.1376~\cite{Aad:2014aqa} \\
$tt$ & $0.86$ & 1506.03062~\cite{Khachatryan:2015sma} &  0.7	 & 1505.07018~\cite{Aad:2015fna}  \\
$Zh$ (pseudo scalar), $\ell\ell b b$ & 0.122 & HIG-15-001~\cite{CMS:2015mba}  &   &  \\
$Zh$ (pseudo scalar), $\ell\ell \tau\tau$ & shape(cut) 0.99(1.26) & HIG-15-001~\cite{CMS:2015mba}  &   &  \\
$hh$, $4b$ & $5.35\times 10^{-2}$ & HIG-14-013~\cite{CMS:2014eda} & $4.2 \times 10^{-2}$  & 1509.04670~\cite{Aad:2015xja}  \\
$hh$, $\gam\gam bb$ & 0.35 & HIG-14-013~\cite{CMS:2014eda} & &  \\
$hh$ $bb\tau\tau$ &  &  & $0.55$  & 1509.04670~\cite{Aad:2015xja} \\
$hh$, combined &  &  & $4.45\times10^{-2}$  & 1509.04670~\cite{Aad:2015xja} \\
$ll$ & $1.45 \times 10^{-3}$ & EXO-12-061~\cite{CMS:2013qca} &  $1.3 \times 10^{-3}$ & 1405.4123~\cite{Aad:2014cka} \\
$Z\gamma$&&&$8.2\times 10^{-3}$&1407.8150 \cite{Aad:2014fha}\\
\hline
\end{tabular}
}

As a reference we also summarize the production cross sections used in this paper for a 750~GeV SM Higgs at 8 and 13~TeV~\cite{higgsxswg}:
\begin{center}
\begin{tabular}{ | c | c | c | c |}
\hline
prod. mode & 8~TeV & 13~TeV & lumi. ratio  \\
\hline
ggF & 0.157~pb & 0.736~pb & 4.693\\
\hline
VBF & 0.05235~pb & 0.1307~pb & 2.496\\
\hline
\end{tabular}
\end{center}

and its branching ratios:

\begin{center}
\begin{tabular}{ | c | c | c | c |}
\hline
$bb$ & $\tau\tau$ & $tt$ & $gg$  \\
\hline
$4.25\times10^{-5}$ & $6.36\times10^{-6}$ & $1.23\times10^{-1}$ & $2.55\times10^{-4}$ \\
\hline
$\gam\gam$ & $Z\gam$ & $WW$ & $ZZ$\\
\hline
 $1.79\times 10^{-7}$ & $1.69\times 10^{-6}$ & $5.86\times10^{-1}$ & $2.90\times10^{-1}$ \\
\hline
\end{tabular}
\end{center}

\section{New states mediating VBF production}
\label{sec:VBF}
\begin{figure}
\includegraphics[width=0.42\textwidth]{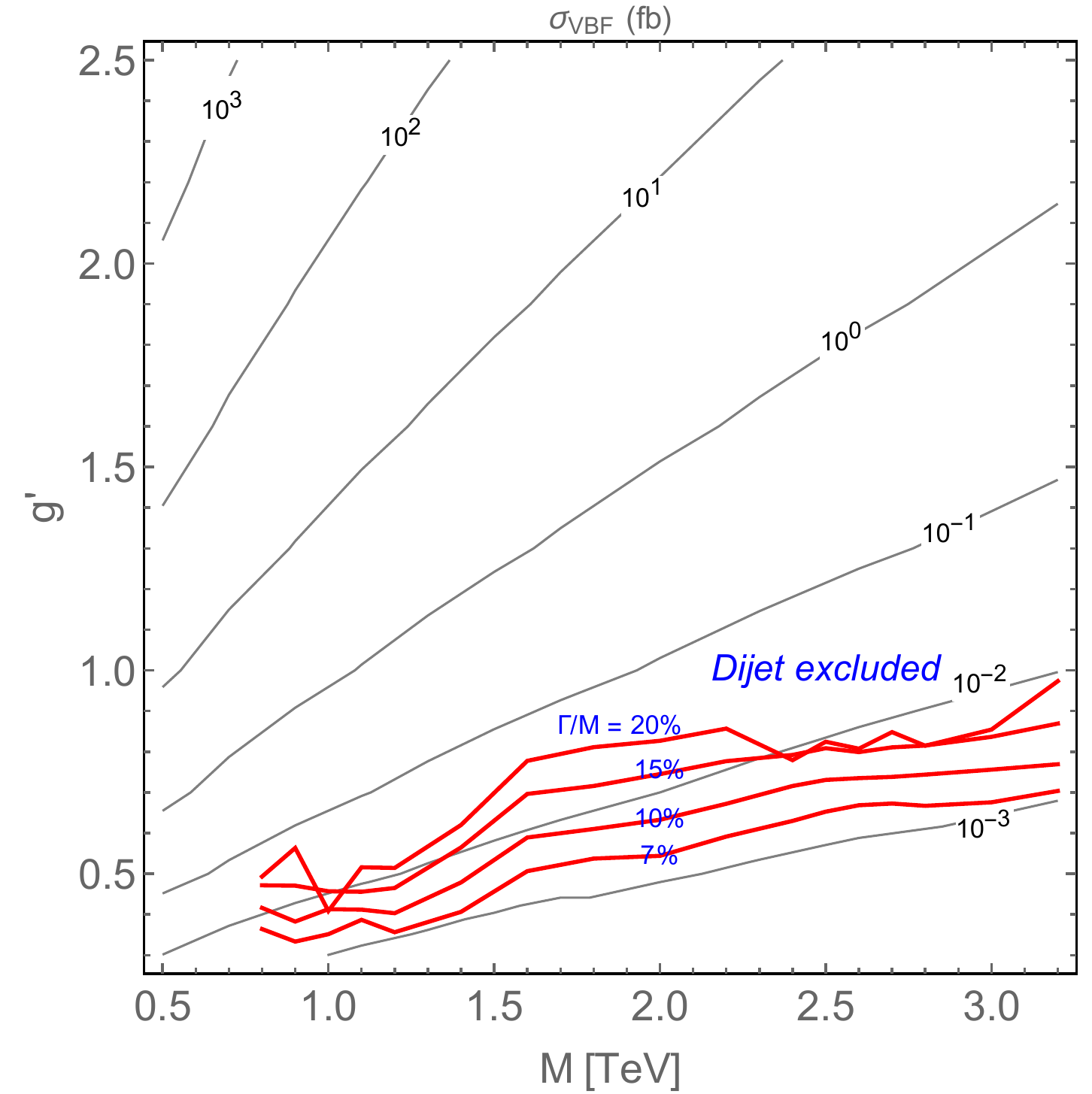}\qquad
\caption{Contours for heavy $W'$ vector boson fusion cross section, $\sig_{VBF}$, in fb. Red solid lines are di-jet exclusion limits on the coupling of a heavy gaussian resonance.\label{figureVBF}}
\label{fig:vbfplots}
\end{figure}

In this appendix we consider the scenario in which the resonance is produced via vector boson fusion (VBF) of new heavy vector states, and we show that di-jet limits exclude the possibility of accounting for the (entire) signal rate. We consider here VBF with a heavy, right-handed $W'$ boson, which is well motivated by $L-R$ symmetric models/extended gauge boson sectors, especially in light of the di-boson anomalies seen in 8 TeV ATLAS and CMS data.   The effective operator is
\be
\mathcal{O}\sim \frac{g'^3}{M_{W'}^3} \, \overline{q}_1\gamma^\mu P_R q_2\,\overline{q}_3\gamma_\mu P_R q_4 \,\Phi
\label{eq:effopVBF}
\ee
where $M_{W'}$ is the mass of the heavy boson, $g'$ is it's coupling and $P_R=(1+\gamma_5)/2$ is the right-handed projection operator. (We nevertheless generate the full kinematics of VBF when using Monte-Carlo to calculate the cross-section.) 

In Fig.~\ref{fig:vbfplots} we plot the cross-section contours of the production of a 750\,GeV $\Phi$ as a function of the mass, $M_{W'}$, and the coupling, $g'$. We observe relatively small production cross sections, which is to be expected since the effective operator eq.~\eqref{eq:effopVBF} is dimension 7. Overlaid are the exclusion limits on the coupling of a heavy gaussian resonance coming from 8 TeV di-jet production, of width/mass, $\Gamma/M_{W'}=$0.07, 0.10, 0.15, and (extrapolated) 0.20. We see that these limits easily exclude the required signal cross-section of $\sim 5-10,\mbox{ fb}^{-1}$. 

VBF production via a heavy generic $V'$ boson ({\it e.g.} a $Z'$ with possibly generic vector and axial couplings) is a separate process and does not interfere with the above; given the simple Lorentz structure of both VBF production of $\Phi$ and of $pp\to V' \to jj$, di-jets will generate similarly strong constraints for this process.

\section{Estimation of the shower shape of a boosted pair of photons}
\label{app:shower}
Here we provide the details on the procedure utilized to estimate the discrepancies between the energy depositions of two photons from a boosted particle of mass $m$, energy $E\gg m$ and proper decay length $c\tau$, from a single prompt photon of energy $E$. We used the shower shape parameterization defined in~\cite{Grindhammer:1993kw}, neglecting fluctuations, to compute the energy deposition into idealized calorimeter cells with the same transverse size as those used by CMS and infinitely long. We considered a $3\times3$ cell array and assumed that either the single photon or the light particle $\phi$ are incident on the center of the array. In the case of $\phi$ decaying to collimated photons, we compute the entrance positions for a given distance and opening angle of the pair and use these positions as starting points of the two showers. After obtaining the energy depositions in the 9 cells for the case of the prompt photon of energy $E$ and the two photons of energy $E/2$ we compute their differences for each cell. We consider a discrepancy only if the energy difference is larger than the single-cell energy resolution quoted by CMS~\cite{Chatrchyan:2013dga}. We then take the value of 2 standard deviations above the mean discrepancy for the set of 9 cells as a proxy of a shower shape difference. For the case of a long lived $\phi$ with fixed energy and proper lifetime, we average the shower shape difference over all the possible decay lengths from the primary vertex to the front of the calorimeter (set at $1\,$m). As a reference point, motivated by the known case of the SM Higgs search,  we compute the quantity defined above for the case of a prompt $\pi^{0}$ of 65~GeV of energy versus a photon of the same energy. The obtained value is used to draw the red/gray boundary in Fig.~\ref{fig:hv}, while we consider the discrepancy for a promptly decaying particle of mass 50~MeV and energy of 65~GeV to draw the white/gray boundary.


\bibliography{refs}

\end{document}